\documentclass[acmtog,nonacm]{acmart}

\acmSubmissionID{176}

\usepackage{booktabs} 

\usepackage{setspace,afterpage,fix-cm,multirow}
\usepackage{bm}
\usepackage{color,graphicx}
\usepackage{colortbl}
\usepackage{enumitem}
\setitemize{noitemsep,topsep=0pt,parsep=0pt,partopsep=0pt,leftmargin=*}
\usepackage{hyperref}
\usepackage[capitalise]{cleveref}
\usepackage[noend]{algpseudocode}
\usepackage{mathtools}
\usepackage[normalem]{ulem}
\usepackage{multirow}
\usepackage{tabstackengine}
\usepackage{caption}
\usepackage{subcaption}
\usepackage{siunitx}
\usepackage{pgfplots,pgfplotstable}
\usepackage{pgf,tikz}
\usepackage{acronym}
\usepackage{listings}
\usepackage{wrapfig}
\usepackage[percent]{overpic}

\definecolor{darkred}{RGB}{180,0,0}
\definecolor{navyblue}{RGB}{9,51,122}
\definecolor{lightblue}{RGB}{240,245,255}
\definecolor{darkblue}{RGB}{40,40,85}
\definecolor{illustratorred}{RGB}{255,170,170}
\definecolor{illustratorblu}{RGB}{55,113,200}

\hypersetup{
  colorlinks=true,
  linkcolor={darkred},
  citecolor={navyblue}
}

\renewcommand{\gg}{\bm{g}}
\newcommand{\nn}{\bm{n}}

\newcommand{\uu}{\bm{u}}

\newcommand{\jj}{\bm{j}}

\makeatletter
\DeclareRobustCommand\onedot{\futurelet\@let@token\@onedot}
\def\@onedot{\ifx\@let@token.\else.\null\fi\xspace}

\acrodef{flip}[FLIP]{Hybrid-Fluid-Implicit-Particle}
\acrodef{g2p2g}[G2P2G]{Grid-to-Particles-to-Grid}
\acrodef{p2g}[P2G]{Particles-to-Grid}
\acrodef{g2p}[G2P]{Grid-to-Particles}
\acrodef{aos}[AoS]{Array-of-Structure}
\acrodef{soa}[SoA]{Structure-of-Array}
\acrodef{aosoa}[AoSoA]{Array-of-Structs-of-Array}
\acrodef{mgss}[MGSS]{Multi-GPU-Static-Spatial-Partition}
\acrodef{mgsp}[MGSP]{Multi-GPU-Static-Particle-Partition}
\acrodef{sph}[SPH]{Smoothed Particle Hydrodynamics}
\acrodef{pbd}[PBD]{Position Based Dynamics}
\acrodef{pic}[PIC]{Particle-In-Cell}
\acrodef{cfl}[CFL]{Courant–Friedrichs–Lewy}
\acrodef{nacc}[NACC]{Non-Associated Cam-Clay}

\usepackage[ruled]{algorithm2e} 

\crefname{algocf}{Alg.}{Algs.}
\Crefname{algocf}{Algorithm}{Algorithms}
\SetAlFnt{\small}
\SetAlCapFnt{\small}
\SetAlCapNameFnt{\small}
\SetAlCapHSkip{0pt}
\IncMargin{-\parindent}

\setcopyright{rightsretained}
\acmJournal{TOG}
\acmYear{2021}\acmVolume{40}\acmNumber{6}\acmArticle{203}\acmMonth{12}
\acmDOI{10.1145/3478513.3480495}

\begin{document}
\title{Ships, Splashes, and Waves on a Vast Ocean}

\author{Libo Huang}
\affiliation{%
\institution{KAUST}
\city{Visual Computing Center, Thuwal 23955, KSA}
 }
 
 \author{Ziyin Qu}
 \affiliation{%
 \institution{UCLA and UPenn}
 }
 
 \author{Xun Tan}
 \affiliation{
 \institution{Zenus Technology}
  \city{Shenzhen}
 }

 \author{Xinxin Zhang}
 \affiliation{
 \institution{Zenus Technology}
  \city{Shenzhen}
 }
 
\author{Dominik L. Michels}
\affiliation{%
\institution{KAUST}
\city{Visual Computing Center, Thuwal 23955, KSA}
}
 \author{Chenfanfu Jiang}
 \affiliation{%
 \institution{UCLA and UPenn}
 }

\renewcommand\shortauthors{Huang et al.}

\begin{abstract}
The simulation of large open water surface is challenging using a uniform volumetric discretization of the Navier-Stokes equations. Simulating water splashes near moving objects, which height field methods for water waves cannot capture, necessitates high resolutions. Such simulations can be carried out using the Fluid-Implicit-Particle (FLIP) method. However, the FLIP method is not efficient for the long-lasting water waves that propagate to long distances, which require sufficient depth for a correct dispersion relationship. This paper presents a new method to tackle this dilemma through an efficient hybridization of volumetric and surface-based advection-projection discretizations. We design a hybrid time-stepping algorithm that combines a FLIP domain and an adaptively remeshed Boundary Element Method (BEM) domain for the incompressible Euler equations. The resulting framework captures the detailed water splashes near moving objects with the FLIP method, and produces convincing water waves with correct dispersion relationships at modest additional costs.


\end{abstract} 

%
%


%
%


\begin{teaserfigure}
    \centering
	\includegraphics[width=\textwidth]{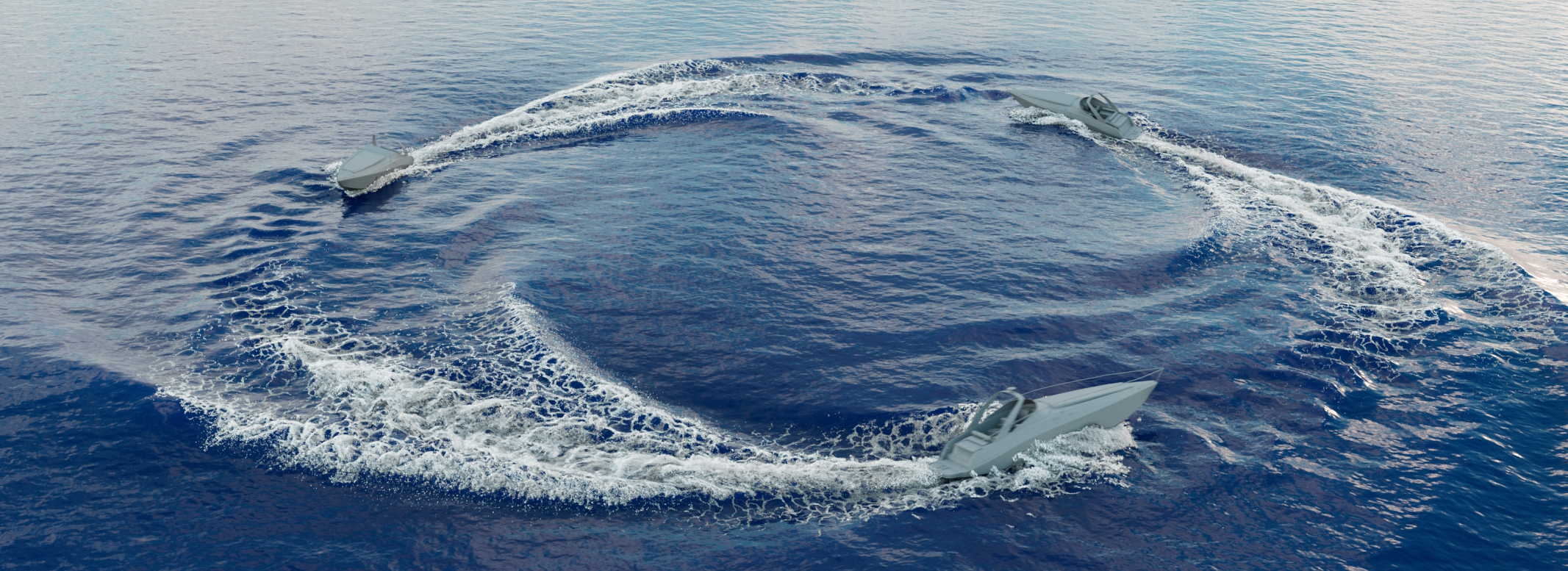}
	\vspace{-5mm}
    \caption{Photorealistic rendering of a simulation comprising three speedboats riding in a circle on the ocean computed with our hybrid approach combining FLIP and BEM.
    \label{fig:teaser}}
\end{teaserfigure}

\maketitle


\section{Introduction} \label{sec:intro}

As the Hong Kong American martial artist and actor Bruce Lee commented, ``water can flow or creep or drip or crash.'' 
From calm waves on a boundless sea to roaring splashes around a speedboat, intriguing large-scale and small-scale aspects of dynamic liquids have attracted great attention in computational physics and computer graphics.  

%
%
Simulating the high-frequency water splashes near the moving structure is relatively easy by a volumetric Navier-Stokes solver. One of such solvers is the popular Fluid-Implicit-Particle (FLIP) method, which was proposed by Brackbill and Ruppel~\shortcite{brackbill1986flip} as an extension to the Particle-In-Cell (PIC) method~\cite{harlow1964particle}, and introduced to computer graphics by Zhu and Bridson~\shortcite{zhu2005animating}. The difficulty is to efficiently propagate the water waves away from the structure.
Simulating the wide and deep water with high resolution FLIP for the low frequency waves seems uneconomic. However, a shallow but wide simulation compromises quality. The depth of the simulated ocean plays a critical role in determining the shape of the surface, as pointed out by Nielsen and Bridson~\shortcite{guidedrobert}. Their approach recovers what was lost in shallow simulations using a pre-computed deep coarse simulation. Along similar lines, adaptive discretizations based on advanced grid structures  \cite{setaluri2014spgrid, ryoichi2020adaptive, ibayashi2018simulating, ando2013highly} have been extensively explored as an option to alleviate the difficulty in computational cost and memory brought by the demanding resolution in volumetric schemes.  However, these adaptive spatial discretizations are known for large dissipation in coarse regions. It is not clear if the long-lasting wide-spreading water waves can be preserved.

Deviating from the adaptivity-related strategies, reduced fluid models represented by height field approaches \cite{jerryocean,waterwavelets,wavepackets, WaveParticles, dispersionwave} have also been widely used in graphics applications for capturing large scale water waves with correct dispersion relationship. Despite their pronounced efficiency for wave phenomena, these methods cannot capture the water splashes near obstacles. It is also non-trivial to two-way couple such approaches with volumetric Navier-Stokes solvers, due to the difficulty in translating back and forth between the height-field information and the velocity-pressure values. Although shallow water equation (SWE) solvers have been coupled with volumetric solvers~\cite{chentanez2015coupling,nilsshallowwatercouple} to produce water splashes, the incorrect dispersion relationship of water waves in deep water can cause unnatural artifacts. 

%
%
Taking a different route, we hybridize a local high-resolution volumetric FLIP discretization and a surface-based BEM ocean discretization for the large, deep ocean. Through a consistent advection-projection discretization of the incompressible Euler equations, we develop a natural way to synchronize velocity-pressure information between the two domains.
With a surface representation of water body, we avoid storing any interior fluid quantities in a large and deep water body, while the surface mesh can be further adapted by criteria such as wave shape and velocity. For water splashes, we adopt a massively parallel FLIP simulation in a small region near the structure and ocean surface to enable intricate close-shot water dynamics.

%
%
Our method is inspired by the following observations. First, the investigations by Da et al.~\shortcite{SurfaceOnlyLiquids} demonstrated the potential of BEM for reproducing complex fluid features with degrees of freedoms only on a surface geometry. Although its demos are at small scale, we speculate that it can produce correct wave dynamics (and it does as shown in the result section) in large scale simulation, and be most efficient compared to volumetric methods for liquid with a large volume/surface ratio such as a large ocean tank. Second, the traditionally challenging aspects of BEM such as solid boundary-handling can be effectively resolved by hybridizing it with a general-purpose volumetric discretization (the FLIP method), making the whole system robust. Finally, the fact that both the BEM and the FLIP method adopt 3D velocity representations of fluid quantities makes the information transfer between them very natural. 

\subsection*{Specific Contributions}

To summarize, we design a new hybrid scheme to efficiently extend water waves generated in a high resolution FLIP simulation. Our method consists of FLIP to BEM coupling where BEM receives correction of volumetric simulation, and BEM to FLIP coupling, where BEM provides the ambient flux of water for the FLIP simulation, similar to the fluxed animated boundary method \cite{FAB}. Our method captures both splashes near structures in multiple moving FLIP domains and long spreading waves on a vast and deep ocean. The water wakes extended by our hybrid approach match oceanography studies \cite{KelvinMachrabaud2013boat}, and brute-force FLIP simulation up to 1.7 billion particles. It also has performance advantages compared to the narrow-band FLIP method \cite{narrowband} and the adaptive octree solver \cite{RyanOctree} implemented in the commercial software Houdini \cite{Houdini}.  

\section{Related Work} \label{sec:related}

\paragraph{Boundary Element Methods in Graphics} The boundary element method (BEM) has a wide range of applications in computer graphics. James and Pai~\shortcite{artdefo} simulated deformable objects interactively with the BEM for linear elasticity. More complex solid behaviors like fracture can also be captured and accelerated through the BEM \cite{brittle}. The BEM is adopted for simulating various fluid phenomena including large scale deep ocean waves \cite{keeler}, fluid chains, water droplets \cite{SurfaceOnlyLiquids}, and even magnetic fluids \cite{SurfaceOnlyFerrofluids}. Besides, BEM has also been adopted in sound rendering~\cite{zheng2010rigid} and geometry processing \cite{justin}.

\paragraph{Hybrid Lagrangian-Eulerian Simulation} Hybrid fluid solvers combining Lagrangian and Eulerian perspectives have gained popularity in recent years. The Particle-In-Cell (PIC) method \cite{PIC}, as a pioneering work of this kind, is limited by its excessive numerical dissipation. The FLIP method~\cite{brackbill1986flip,zhu2005animating} effectively improves this issue by only interpolating grid changes back to particles. More recent approaches such as the APIC method \cite{apic} and the PolyPIC method \cite{polypic} were proposed to reduce noise and instability of the FLIP method while keeping numerical dissipation during particle grid transfer under control. By coupling an Eulerian fluid solver with a narrow band of FLIP particles, Ferstl et al.~\shortcite{narrowband} greatly reduced the particle counts while maintaining an indistinguishable result from a pure FLIP simulation. Losasso et al.~\shortcite{losassoSPH} incorporated Smoothed Particle Hydrodynamics (SPH) within a grid-based simulation. Domain decomposition approaches \cite{golas2012large} effectively reduce the memory and computational costs by combining Eulerian simulations with vortex particles. Recently, Yue et al. \shortcite{yue2018hybrid} coupled material point and discrete element methods to simulate granular material.

\paragraph{2D Methods} SWE methods have been adopted due to their simplicity. Thuerey et al.~\shortcite{nilsrealtimeoverturning} enhanced this model with the capability of handling overturning waves. Solenthaler et al. \shortcite{SPHshallow} used SPH particles to the SWE. Procedural waves \cite{jerryocean} are effective on simulating vast oceans, however it is non-trivial to handle boundaries. Procedural waves could be enhanced by satisfying solid boundary conditions \cite{procedualboundaryaware}, handling moving obstacles with wavelet discretization \cite{waterwavelets}, handling static obstacles \cite{waterwavewavefront}, or adding high frequency details while matching the given coarse waves \cite{robertheightfield}. Schreck et al.~\shortcite{fundmentalwaterwave} simulated water waves through fundamental solutions which do not require grids or meshes. However, in their boat wake scene, the wake does not show the feather-like pattern in reality. Lagrangian wave particles simulate water waves with propagating particles \cite{WaveParticles, wavepackets}. Canabal et al. \shortcite{dispersionwave} applied pyramids of convolution kernels with finite support to 2D height fields to get realistic wave dispersion relationships. The amplitude of the wave animation is given as a parameter tuned by artists. 2D methods are limited to only capture waves. 

\paragraph{2D-3D Coupled Methods} To capture water splashes, coupling 2D methods with 3D methods is a natural choice. Chentanez et al.  \shortcite{chentanez2015coupling} coupled particles, Eulerian grid solver, and SWE. Thuerey et al. \shortcite{nilsshallowwatercouple} coupled the SWE with a 3D solver employing a Lattice Boltzmann Method (LBM). However, as commented by Chentanez et al. ``the SWE and 3D
grids simulate inherently different physical phenomena''. The waveform and propagation speed is visibly different from that of a 3D solver.

\paragraph{Spatial Adaptivity} Fluid simulations in large domains such as oceans could be troublesome with uniform volumetric solvers due to the high computational costs and limited memory capacities.  Adaptive methods such as octrees \cite{losassooctree, ryoichi2020adaptive, hong2009adaptive, aanjaneya2017power} are usually dissipative in coarse regions. It is not sure if the water waves could be preserved away from the fine region. Tall cell methods \cite{tallcell,chentanez2011real} were initially designed for shallow water regime. As reported by Irving et al. \shortcite{tallcell}, the performance gain due to use of tall cells is only about a factor of two compared to uniform methods. It offers no adaptivity along horizontal directions, so simulating wide water bodies is still challenging for tall-cells methods. Chimera grid methods \cite{chimera} require sophisticated mesh handling, and suffer from inefficient
pressure projection. The far-field grid method \cite{bofarfield} is suitable for propagating waves from a single domain, however, supporting multiple moving domains is difficult. Nielsen and Bridson~\shortcite{tiletreeadaptive} adopted adaptive tile tree for FLIP simulation. Goldade et al.~\shortcite{RyanOctree} incorporated an adaptive octree for viscosity and pressure projection.

\paragraph{Guided Simulation} Nielsen and Bridson~\shortcite{guidedrobert} proposed guided shapes for high resolution simulations to reduce the design cycles. Later, Nielsen et al.~\shortcite{localizedguided} further improved this method by localizing the guided shapes only requiring the surface mesh. Bojsen-Hansen and Wojtan~\shortcite{nonreflecting} proposed a way for achieving generalized non-reflecting boundary conditions for localized re-simulation. Absorbing boundary conditions has also been investigated \cite{PML2010soderstrom}. Similar to our approach, flux boundary method \cite{FAB} coupled outer prescribed ocean motion with FLIP/MPM solver through boundary flux, while it is non-trivial to achieve two way coupling with their method.

\paragraph{Lagrangian Methods} Lagrangian particles such as SPH are also adopted for fluid simulation \cite{ihmsen2013implicit,Huang:2019:ALS:3306346.3322973}. Secondary effects such bubble, foam, and sprays \cite{sprayfoambubbles} can visually enhance the look of liquid in post processing. Recently, adaptivity has also been investigated in the context of SPH particles \cite{InfiniteContinuousAdaptiveSPH}.

\paragraph{Engineering Paradigms} Early work from Longuet-Higgins and Cokelet~\shortcite{longuet1976deformation} utilized the BEM to numerically compute steep water waves. Higher order time-stepping techniques with BEM are essential for resolving more general wave phenomena \cite{GRILLI198997}. The BEM could also be combined with the Finite Element Method (FEM) and be applied for analyzing the nonlinear interactions between waves and bodies \cite{WU2003387}. More applications and experiments like propeller blades \cite{propeller}, added resistance of boats in waves \cite{CHEN2018112}, and bubble dynamics \cite{ZHANG2015208} have been effectively conducted using the BEM.

\section{Method}

\begin{figure}[h]
    \centering
   \includegraphics[width=0.47\textwidth]{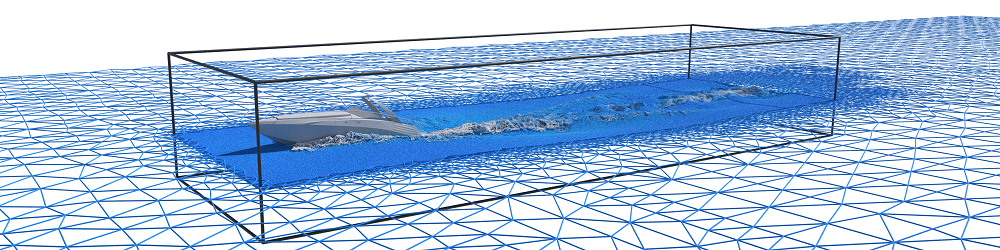}
    \caption{A moving cage follows the movement of a boat on the ocean. The large body of water is solved approximately with the BEM (only top layer rendered), and the resulting water flow enables high resolution simulation with the FLIP method inside the cage.}
    \label{fig:method:iron_cage_on_ocean}
\end{figure}


\noindent
The core idea can be intuitively explained by the following imaginary scene: a boat is moving on the ocean, while there is a moving cage containing the boat, water, and air as illustrated in Figure~ \ref{fig:method:iron_cage_on_ocean}. We use a large closed mesh to represent the water of the ocean. Initially there were six faces. The top face is free surface while other five faces are kinematic solid boundaries. The six faces are further tessellated into triangles for the use of BEM simulation \cite{SurfaceOnlyLiquids}. In Figure~ \ref{fig:method:iron_cage_on_ocean}, we only render the top triangles to avoid visual distraction. The FLIP method is used exclusively inside the cage. Liquid-boat interactions are handled by the FLIP method, and the BEM vertices inside the cage follow the motion of the FLIP liquid surface. The BEM in return provides the water flux below free surface on the cage boundary for the FLIP simulation. FLIP influences the BEM only by changing its vertex velocity, and the BEM influences FLIP only by providing the flux boundary condition below water. It is similar to the FAB method \cite{FAB}, except the flux boundary condition in the FAB method is prescribed by animators, while the flux boundary condition here is provided by the BEM, which is not prescribed but driven by the FLIP simulation. This allows waves generated by solids to propagate out of the FLIP domain, and even to another FLIP domain via the BEM mesh.

We briefly cover both FLIP and BEM, and then explain our main contribution, the coupling workflow between them. The entire workflow is summarized in Algorithm \ref{algo:One_Frame} and illustrated in Figure \ref{fig:flow_chart}.

\subsection{FLIP and BEM} \label{sec:eqnsystem}
The governing equation system of our work is the incompressible Euler equation
\cite{bridson2007fluid}:
\begin{eqnarray}
\frac{D \uu}{D t} &=& -\frac{1}{\rho}\nabla P + \gg \label{eq:euler1}\,,\\
\nabla \cdot \uu &=& 0\label{eq:euler2}\,,
\end{eqnarray}
where $D/D t := \partial/\partial t+\uu\cdot\nabla$ is the material derivative, $\uu$ is the velocity field, $\rho$ is the constant liquid density, $P$ is the pressure, and $\gg$ is the gravitational acceleration. 

In both, BEM and FLIP, the equation is solved by operator splitting in an advection-projection style. Assume, in the beginning of the time step, the velocity is $\uu^-=\uu(t)$. After the advection, the velocity becomes $\uu$. Finally, after the projection, the velocity becomes $\uu^+=\uu(t+\Delta t)$, where $\Delta t$ is the time step size.

In the advection step, we solve the advection equation
\begin{equation}
    \label{eq:adv}
    \frac{D \uu}{D t}=0
\end{equation}
to obtain $\uu$ from $\uu^-$, and then add gravity to it.
In the pressure projection step, the incompressibility constraint $\nabla\cdot\uu=0$ is enforced by solving ${\partial\uu}/{\partial t} = -{\rho}^{-1}\nabla P$. The associated Poisson's equation is
\begin{eqnarray}
\nabla^2 P &=& \frac{\rho}{\Delta t}\nabla\cdot\uu\,,\nonumber\\
P|_{\Gamma_\text{D}} &=& P_{\text{bc}}\,,\label{eq:pressure_poisson}\\
\frac{d P}{d n}|_{\Gamma_\text{N}}&=&\frac{\rho}{\Delta t}(\uu-\uu_{\text{solid}})\cdot \nn\,,\nonumber
\end{eqnarray}
where $\Gamma_\text{D}$ and $\Gamma_\text{N}$ are liquid-air (Dirichlet) and liquid-solid (Neumann) boundaries respectively. $P_{\text{bc}}$ is the Dirichlet pressure boundary condition which usually consists of surface tension, $\uu_{\text{solid}}$ is the desired liquid velocity on the solid boundary, which can introduce inflow and outflow, and $\nn$ is the external normal vector of the liquid domain.
Eq.~\eqref{eq:pressure_poisson} is solved to get the pressure $P$, and the velocity is then updated by
\begin{equation}
\label{eq:projection_update}
    \uu(t+\Delta t) = \uu^+ =  \uu - \frac{\Delta t}{\rho}\nabla P\,.
\end{equation}

The pressure projection Eq.~\eqref{eq:pressure_poisson} is solved differently in the BEM and FLIP. 
For the FLIP simulation, we generally follow Zhu and Bridson \shortcite{zhu2005animating}. In the pressure projection step, we use the approach of Ng et al. \shortcite{Ng09Boundary} to handle the liquid-solid boundary, and refer to Gibou et al. \shortcite{gibou2002second} for the free-surface boundary condition. Since we focus on large scenes, we neglect the surface tension.

In short, the BEM simulates the motion of liquid by evolving triangle meshes. Each vertex carries position and velocity information of the liquid surface. In the advection step, each vertex moves with corresponding velocity to the new position. However, when these vertices move, collisions between points to triangles, and edges to edges happen. Furthermore, when two droplets collide, topology changes are required to join two disjoint droplet meshes together. Such collision handling and topology changes are handled by a mesh-based surface tracking program LosTopos \cite{LosTopos}. The input of LosTopos is the original vertex positions, triangle connectivity, and the target vertex positions. It resolves collisions, topology changes, and outputs a new set of vertex positions and triangle connectivity. The output mesh is intersection-free and watertight. After the advection, the surface velocity field may not represent the boundary value of a divergence-free velocity field. Therefore, it is followed by a projection step consisting of Helmholtz decomposition (H.D.) and solving a Laplace equation with mixed boundary condition. We refer the reader to the work of Da et al. \shortcite{SurfaceOnlyLiquids}, and the work of Huang and Michels \shortcite{SurfaceOnlyFerrofluids} for details. We prefer the H.D. in the implementation of Da et al. \shortcite{SurfaceOnlyLiquids} because the look is closer to our reference 3D simulation, see Figure \ref{fig:res:BEM_VS_FLIP_crownsplash:BEM} and \ref{fig:res:BEM_VS_FLIP_crownsplash:FULLHDBEM}. The theoretical analysis of such preference is presented in the supplemental material.

\begin{algorithm}[t]
\SetKwData{Left}{left}\SetKwData{This}{this}\SetKwData{Up}{up}
\SetKwFunction{FLIPtoBEMCoupling}{FLIPtoBEMCoupling}
\SetKwFunction{BEMAdvection}{BEMAdvection}
\SetKwFunction{BEMProjection}{BEMProjection}
\SetKwFunction{FLIPAdvection}{FLIPAdvection}
\SetKwFunction{BEMtoFLIPCoupling}{BEMtoFLIPCoupling}
\SetKwFunction{FLIPProjection}{FLIPProjection}
\SetKwFunction{Max}{max}
\SetKwFunction{Min}{min}
{\bf Input:} FLIP particles and BEM mesh at time $t$.\\
{\bf Output:} FLIP particles and BEM mesh at time $t+\Delta t_f$.
\BlankLine
$t_\text{BEM}\leftarrow 0$\;
\While{$t_\text{BEM}<\Delta t_\text{frame}$}{
$\Delta t_\text{BEM}\leftarrow$ \Min{$\Delta x_\text{BEM}$ / \Max{$v_\text{BEM}$}, $\Delta t_\text{frame} - t_\text{BEM}$}\;
$ t_\text{BEM}\leftarrow t_\text{BEM} + \Delta t_\text{BEM}$\;
\tcc{Figure \ref{fig:flow_chart}, Plot (2), Figure \ref{fig:method:BEMcorrection} (left), Section \ref{sec:method:FLIP2BEM}}
\textcolor{blue}{\FLIPtoBEMCoupling{$\Delta t_\text{BEM}$, \text{FLIP SDF and velocity}}}\;
\BEMAdvection{$\Delta t_\text{BEM}$}\;
\tcc{Figure \ref{fig:flow_chart}, Plot (3)}
\BEMProjection{$\Delta t_\text{BEM}$}\;
}

$t_\text{FLIP}\leftarrow 0$\;
\While{$ t_\text{FLIP} < \Delta t_\text{frame}$}{
$\Delta t_\text{CFL} =$ $ \text{cfl}()\Delta x_\text{FLIP}$\ / \Max{$v_\text{FLIP}$}\;
$\Delta t_\text{FLIP}\leftarrow$ \Min{$\Delta t_\text{CFL}$, $\Delta t_\text{frame} - t_\text{FLIP}$}\;
$ t_\text{FLIP}\leftarrow t_\text{FLIP} + \Delta t_\text{FLIP}$\;
\tcc{Figure \ref{fig:flow_chart}, Plot (4)}
\FLIPAdvection{$\Delta t_\text{FLIP}$}\;
\tcc{Figure \ref{fig:method:BEMcorrection} (right), Section \ref{sec:method:BEM2FLIP}}
\textcolor{blue}{\BEMtoFLIPCoupling{\text{BEM mesh and velocity}}}\;
\tcc{Figure \ref{fig:flow_chart}, Plot (5)}
\FLIPProjection{$\Delta t_\text{FLIP}$}\;
}

\caption{A single frame is simulated using our coupled BEM and FLIP approach. A time step size of $\Delta t_\text{frame}$ is used, $\Delta x_\text{BEM}$ denotes the minimum edge length of the BEM mesh, $v_\text{BEM}$ indicates the BEM vertex velocity, $\Delta x_\text{FLIP}$ denotes the FLIP voxel size, and the FLIP velocity is given by $v_\text{FLIP}$. The CFL number is given by cfl(). Extra coupling steps are written in \textcolor{blue}{blue}.}
\label{algo:One_Frame}
\end{algorithm}\DecMargin{1em}

\begin{figure*}
    \centering
    \begin{overpic}[width=0.98\textwidth]{graphics_final/whole_flow_chart_notext.pdf}
    \put(0,30){(1)}
    \put(30,30){BEM}
    \put(34.5,30){(2)}
    \put(42,30){Guided Advection}
    \put(69,30){(3)}
    \put(80,30){Projection}
    \put(30,14){FLIP}
    \put(34.5,14){(4)}
    \put(45,14){Advection}
    \put(69,14){(5)}
    \put(76,14){Projection with Flow}
    \put(7,3){Next frame FLIP boundary}
    \put(7,4.73){FLIP boundary}
    \put(7,6.45){BEM mesh}
    \put(7,8.17){Interpolated BEM velocity}
    \put(7,9.89){FLIP velocity}
    \put(7,11.61){BEM velocity}
    \end{overpic}
    \caption{Flow chart of a single time step. (1) In this scene a water ball impacts the surface causing it to bend inward. The light blue region is the FLIP liquid domain. The solid dark blue and red lines are the boundary of the FLIP liquid and the BEM surface mesh respectively. The dashed blue line indicates the FLIP free surface after advection. The dark blue and red arrows indicate the liquid velocity on the boundary of the FLIP and BEM simulation respectively. The black arrows are the FLIP boundary velocity interpolated from the BEM mesh. (2) BEM vertices move to its new positions guided by the FLIP free surface and velocity; see Section \ref{sec:method:FLIP2BEM} and Figure \ref{fig:method:BEMcorrection} (left). (3) After the BEM projection step, the velocity field becomes divergence free again. (4) FLIP advection. (5) The boundary velocity of the FLIP domain is interpolated from the BEM surface velocity field. It is then plugged in the projection step of FLIP to make the velocity field divergence free; see Section \ref{sec:method:BEM2FLIP}. This concludes a single time step.}
    \label{fig:flow_chart}
\end{figure*}

\subsection{FLIP Influence on the BEM}
\label{sec:method:FLIP2BEM}

\begin{figure}
\begin{overpic}[width=0.47\textwidth]{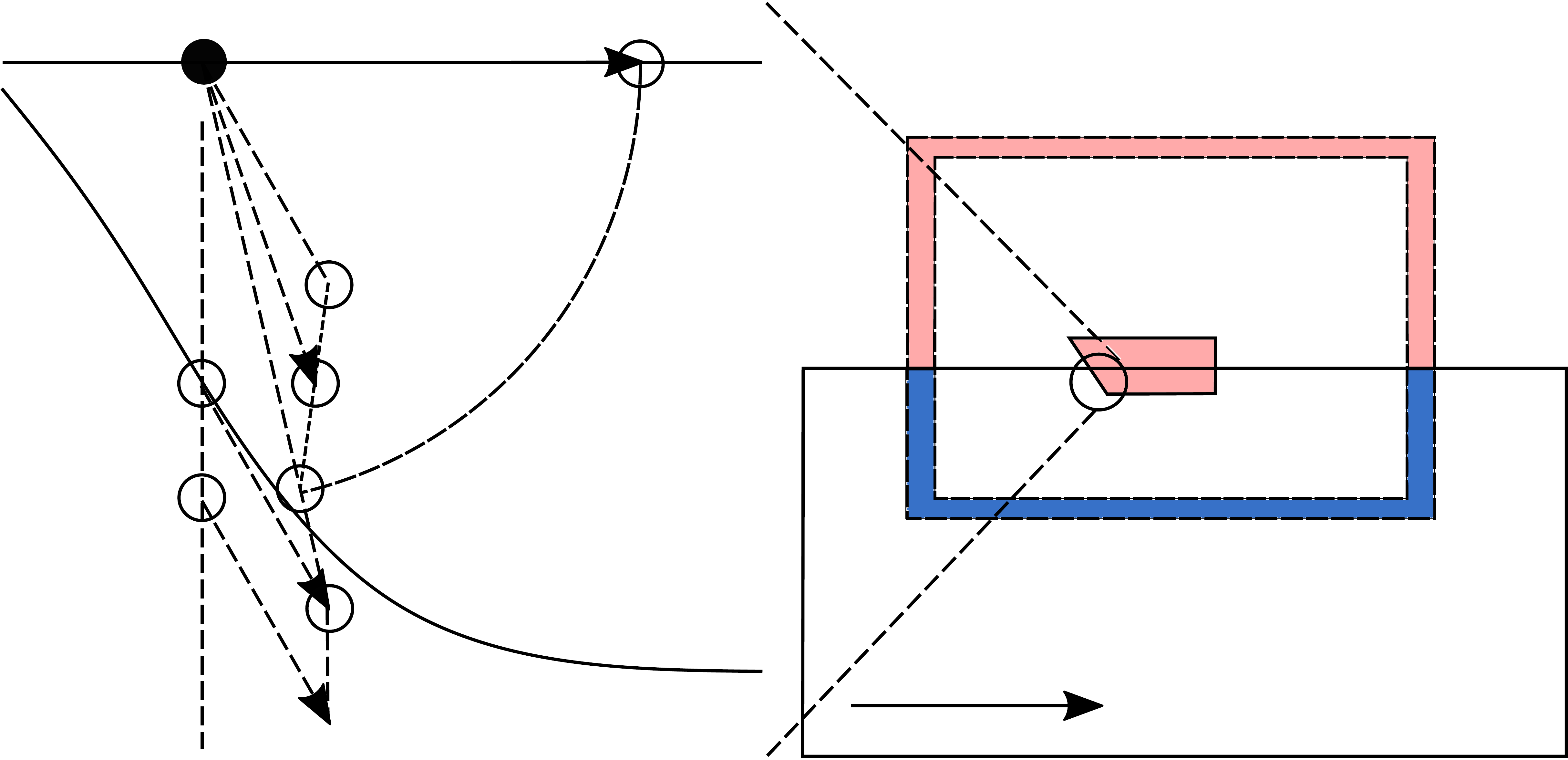}
\put(10,47){BEM vertex}
\put(9,41){$x_0$}
\put(8,23){$x_1$}
\put(22,6){$x_2$}
\put(22,1){$v_\text{FLIP}$}
\put(36,41){$x_3$}
\put(21,15){$x_4$}
\put(23,31){$x_5$}
\put(22,23){$x_\text{final}$}
\put(2,14){$x_\text{sample}$}
\put(30,10){FLIP surface}
\put(58,5){Flow}
\put(75,5){BEM domain}
\put(68,11){\textcolor{illustratorblu}{Fill zone}}
\put(70,28){Boat}
\put(65,34){FLIP domain}
\put(63,41){\textcolor{illustratorred}{Sink zone}}
\end{overpic}
\caption{Illustration of the advection step in the coupling mechanism between BEM and FLIP. The details are explained in Section~\ref{sec:method:FLIP2BEM} and~\ref{sec:method:BEM2FLIP} . Left: A BEM vertex at $x_0$ moves under the guidance of FLIP to its final destination at $x_\text{final}$.  Right: A boat is moving on the ocean represented by the BEM domain. A buffer region around the FILP domain is divided into the sink zone where  particles are removed, and the fill zone where new particles are generated.}
\label{fig:method:BEMcorrection}
\end{figure}

\begin{figure}
\includegraphics[width=0.47\textwidth,trim={0 0 0 4cm},clip]{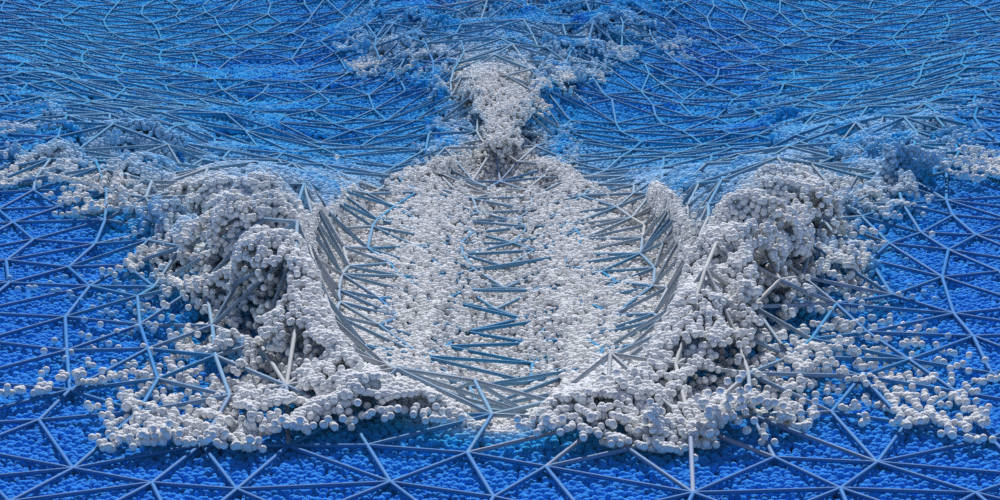}
\includegraphics[width=0.47\textwidth,trim={0 0 0 4cm},clip]{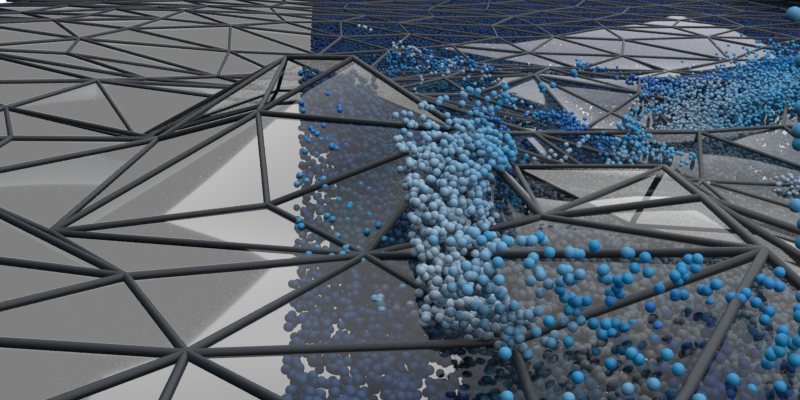}
\caption{Top: The BEM mesh moves under the guidance of FLIP particles, closely matching its surface. The incoming boat is removed for better visualization of the bottom FLIP surface. With reference coordinate system fixed at the boat, white color indicates slow velocity while blue indicates high velocity. Bottom: Particles move to the left, exiting the FLIP domain. Particles in the buffer zone and above the BEM mesh are removed, while particles below the BEM mesh are filled to keep constant density. }
\label{fig:method:kill_fill_embedding}
\end{figure}

This section corresponds to the guided advection step in Figure \ref{fig:flow_chart}.
The BEM takes information from the FLIP simulation by tracking the water flow on the liquid-air and liquid-solid interfaces of the FLIP system. However, just advecting BEM vertices with FLIP velocity field may leads to a drift of free surface. Therefore we must consider both the flow velocity and liquid interface.

We expect the FLIP to BEM coupling to be non-intrusive to the existing BEM program, so we decide to achieve the goal by modifying the velocity of each vertex before the BEM time stepping function, instead of directly matching the interface position and flow velocity separately inside the advection function of the BEM.

There are splashes in the FLIP simulation. Letting the BEM track all these detailed droplets is unnecessary because those droplets will eventually merge into the main body water. When they merge, the BEM is going to track the main surface again. Therefore, we let the BEM track a smoothed FLIP liquid signed distance function (SDF).

The detailed algorithm is as follows. The key steps are illustrated in Figure \ref{fig:method:BEMcorrection}. Snapshots from actual simulations are shown in Figure \ref{fig:method:kill_fill_embedding}. We use the smoothed SDF of the FLIP liquid and velocity field at the beginning of current frame as a guidance to find the target position that a BEM vertex is supposed to be at the end of a substep. Based on that we modify the velocity.

First, given the current BEM vertex position $x_0$, we search from a point below $x_0$ and inside the FLIP fluid, stepping $0.5\Delta x_\text{FLIP}$ upward until the sampled liquid SDF becomes positive, then based on the sampled positive SDF we move downward to get the sub-voxel position of the interface $x_1$. If there is no intersection $x_1$, then all following operations are skipped.

In the current cut-cell formulation of Neumann boundary conditions in the FLIP \cite{Ng09Boundary}, we find it helpful to produce water splashes in front of the boat if we mix the liquid velocity with solid velocity proportional to the solid fraction of the corresponding voxel face. For example, if the voxel face of a velocity component has $30\%$ of solid, with an artificial friction coefficient of 0.8, and the boat is static, the blended component velocity would be $76\%$ of the original value. However, the side effect is that BEM takes such low velocity as a sign of large resistance, producing exaggerated wake amplitudes (Figure \ref{fig:res:wake_pattern_depth}). To alleviate this side effect, we sample the FLIP velocity not directly at the surface point $x_1$, but a few voxels below:
\begin{eqnarray}
    x_\text{sample} &=& x_1 - (0,\alpha \Delta x_\text{FLIP},0)\,, \label{eq:sample_lower}\\
    v_\text{FLIP} & = & \text{VelocitySample}(\text{FLIP},x_\text{sample})\,.
\end{eqnarray}
The BEM still tracks the correct free surface, but feels less resistance due to faster flow velocity sampled from the FLIP. The effects of sampling depth $\alpha$ will be investigated in Section \ref{sec:speedboat} and in Figure \ref{fig:res:wake_pattern_depth}.

The sampled velocity $v_\text{FLIP}$ determines the ideal target location of the BEM vertex $x_2$ at the end of the frame:
\begin{equation}
    x_2 = x_1 + \Delta t_\text{frame}v_\text{FLIP}\,.
\end{equation}
A problem with setting $x_2$ as the substep target position for the BEM vertex is that it would attempt to have a high velocity to travel very long distance ($x_2-x_0$) in a short period of time if there is initial surface mismatch or substep time $\Delta t_\text{BEM}$ is smaller than frame time $\Delta t_\text{frame}$. Therefore, we limit the distance that a BEM vertex can travel by its original velocity $v_\text{BEM}$ and substep time $\Delta t_\text{BEM}$ to get the limited surface target $x_4$:
\begin{eqnarray}
    x_3 &=& x_0 + \Delta t_\text{BEM}v_\text{BEM}\,,\\
    x_4 & = & x_0 + \frac{x_2-x_0}{|x_2-x_0|}\text{min}(|x_2-x_0|,|x_3-x_0|)\,.
\end{eqnarray}
However, if all BEM vertices are initially static, they would not move at all. Therefore we need to blend the velocity with the FLIP flow velocity $v_\text{FLIP}$. The target position with pure FLIP flow velocity is $x_5$:
\begin{equation}
    x_5 = x_0 + \Delta t_\text{BEM}v_\text{FLIP}\,.
\end{equation}
The final substep target position of the BEM vertex is a linear mixture of free surface target $x_4$ and FLIP flow target $x_5$. We just take their mid point:
\begin{eqnarray}
    x_\text{final} &=& \frac{1}{2}(x_4 + x_5)\,.
\end{eqnarray}
Finally, the vertex velocity is obtained:
\begin{equation}
    v_\text{BEM} = \frac{x_\text{final}-x_0}{\Delta t_\text{BEM}}\,.
\end{equation}
Although the SDF and velocity of the FLIP are taken at the beginning of the frame, the surface target position $x_2$ is at the end of the frame as an approximation of where the FLIP surface is supposed to be. An alternative way is to advect the SDF and velocity to the end of BEM substep time, and design another workflow, but it is more costly. When it is not the first substep, the projection from $x_0$ to $x_1$ actually do not happen at the beginning of the frame. To force it to happen at the frame beginning, one can estimate where the vertex would be by backtracking with the vertex velocity backward to the frame beginning, find the vertical projection point $x_1$, and estimate the target position $x_2$ at the end of the frame. However, we did not observe visual differences in this more intricate approach. Also most of the time there is only one BEM substep.

\subsection{BEM Influence on the FLIP}
\label{sec:method:BEM2FLIP}

The BEM provides the water flow boundary condition of the FLIP domain. The difficulty is to know the velocity inside the BEM mesh. If this interpolation problem is to be solved by potential flow or Stokes flow \cite{bhattacharya2012steady}, the whole BEM domain would be voxelized and equations are inefficiently solved. Finding the nearest point velocity on the surface seems to be the simple solution, but the velocity field is not physical, see Figure \ref{fig:res:Stokes_wave}. Therefore we propose to use boundary integral over BEM triangles to get the velocity on the FLIP Neumann boundary inside BEM mesh:
\begin{equation}
\uu(x)=\nabla\int_\Gamma\frac{\nn(y)\cdot\uu(y)}{4\pi|x-y|}d s_y -\nabla\times\int_\Gamma\frac{\nn(y)\times\uu(y)}{4\pi|x-y|}d s_y\,,\label{eq:u_HD}
\end{equation}
where $x$ is inside the BEM domain, and $y$ is on the BEM boundary $\Gamma$. If we introduce two auxiliary variables $c$ and $\jj$ (similar to charges and currents in electrodynamics), the above equation becomes:
\begin{eqnarray}
    c(y)&=&\nn(y)\cdot\uu(y)\,,\\
    \jj(y)&=&\nn(y)\times\uu(y)\,,\\
    \uu(x)&=&\nabla\int_\Gamma\frac{c(y)}{4\pi|x-y|}d s_y -\nabla\times\int_\Gamma\frac{\jj(y)}{4\pi|x-y|}d s_y\,,\\
    &=&\nabla\int_\Gamma\frac{c(y)}{4\pi|x-y|}d s_y +\int_\Gamma\jj(y)\times\nabla\frac{1}{4\pi|x-y|}d s_y\,.
\end{eqnarray}

Since $x$ is not on the boundary $\Gamma$, $|x-y|\neq0$. The above integral is not singular. The integral is calculated as the summation of integrals of each triangle on the BEM mesh. On each triangle, the normal direction $\nn$ is constant and the velocity $\uu$ is the barycentric interpolation of vertex values. Accordingly $c(y)$ and $\jj(y)$ on the triangle are calculated based on the interpolated velocity. On each triangle, we select three symmetric quadrature points, whose barycentric coordinates are $2/3, 1/6, 1/6$ with rotation, and weights are just one third of the triangle area. 

The interpolation point $x$ can still be close to the quadrature point which introduces error. The easy solution is to use nearest neighbor interpolation, or one can apply the more expensive analytic integration of the fundamental
solution $1/|x-y|$ over
\begin{wrapfigure}{r}{0.15\textwidth} 
    \includegraphics[width=0.15\textwidth]{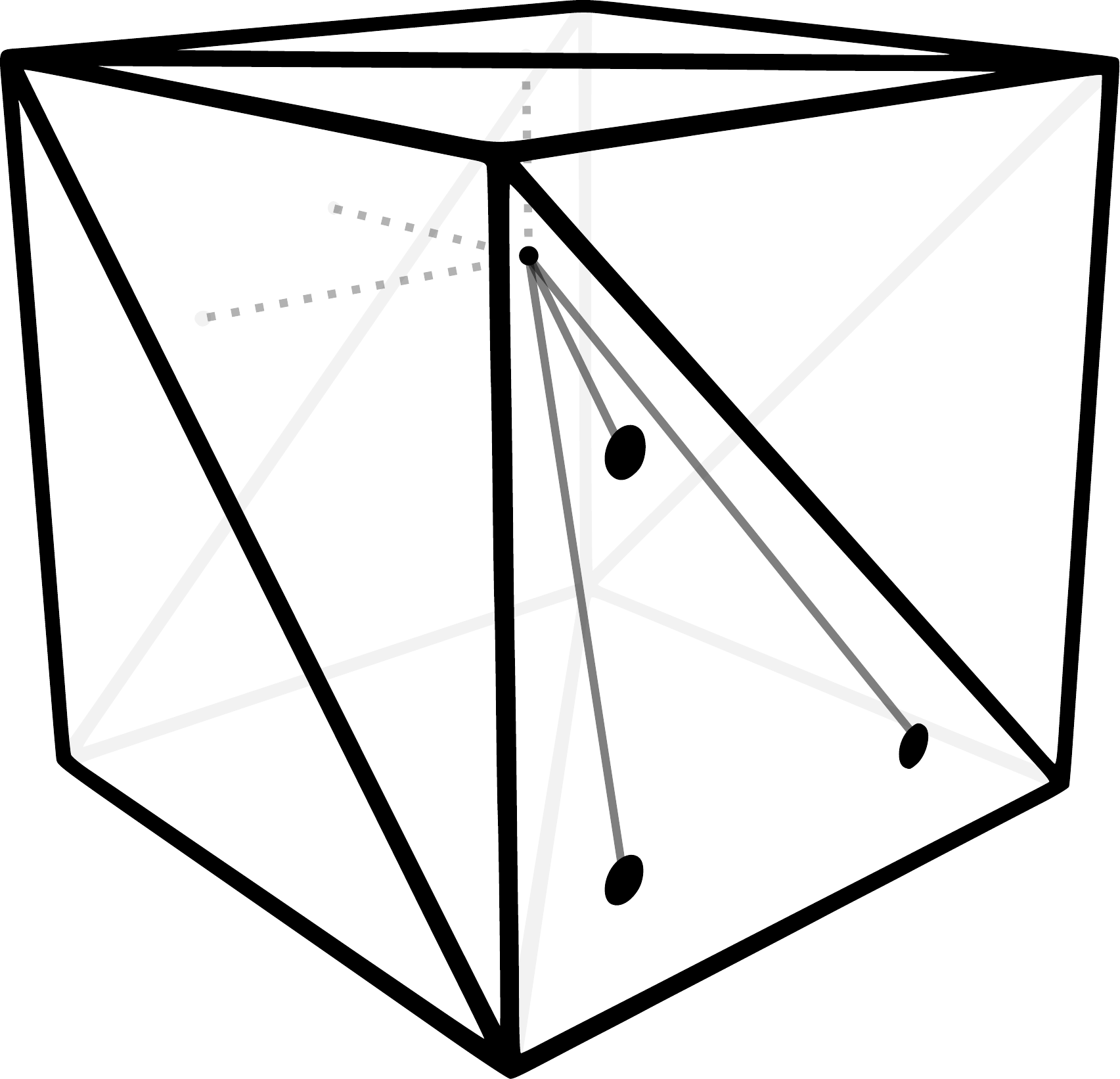}
\end{wrapfigure}
triangles, as adopted by Huang and Michels \shortcite{SurfaceOnlyFerrofluids}. 
The brute force integration for each position on the FLIP boundary is expensive. We take the advantage that the interpolated velocity field inside BEM mesh is smooth. We create a narrow band velocity volume which typically has grid resolution of $4\Delta x_\text{FLIP}$ around the FLIP boundary. The interpolated velocity based on boundary integral is zero if $x$ is outside the BEM mesh. Therefore when creating this narrow band velocity volume, we only sample velocity inside the BEM mesh, and extrapolate one layer to enhance coverage. Otherwise the velocity near BEM free surface would be affected by the zero velocity sample outside.
In addition to boundary velocity, we also need to add and remove particles. Otherwise when the water flows in, FLIP particles left the influx Neumann boundary, and there is a gap between particles and Neumann boundary. Therefore we just need to fill the gap. The gap size is related to the CFL condition, so it is usually a few voxels after advection. Inside the Neumann boundary, we set a fill zone, where particles are filled. Since our maximum CFL is 4, we use 4 voxels wide fill zone. A voxel in the fill zone and inside the BEM mesh are filled with particles, whose initial velocity is given by the narrow band velocity volume obtained from boundary integral. We also set a sink zone where particles are removed. The sink zone is similarly defined except it is outside of the BEM mesh. See Figure \ref{fig:method:BEMcorrection} and  \ref{fig:method:kill_fill_embedding}.

\section{Implementation}
\label{sec:implementation}
The BEM part is based on the open source code of Da et al.~\shortcite{SurfaceOnlyLiquidsCode} with CUDA acceleration provided by Huang and Michels~\shortcite{SurfaceOnlyFerrofluids}. The FLIP code is based on the open source implementation of Batty et al.~\shortcite{variationalFLIP} and Ng et al.~\shortcite{Ng09Boundary} provided by Batty~\shortcite{FLIPcode}. However, we re-wrote the whole FLIP simulation with OpenVDB~\cite{museth2013vdb} to support sparse data structures.

During the guided advection step in the BEM, the liquid SDF of FLIP is represented as an OpenVDB float grid. We first re-initialize the liquid SDF in FLIP simulation with OpenVDB function ``levelSetRebuild'', since only in the one-voxel neighborhood of a particle we have a valid SDF, while the following Gaussian filter needs wider bandwidth. We then apply three passes of one-voxel-wide Gaussian filters provided by OpenVDB to smooth the liquid SDF. As illustrated in Figure~\ref{fig:method:BEMcorrection} (left), we need to find the BEM vertex vertical projection point on the FLIP surface. We search from the bottom of the FLIP domain, and step upward until the sampled liquid SDF becomes positive. The search step size is half of the FLIP voxel size. The projection step of the BEM is a Galerkin boundary element method as described and implemented by Huang and Michels~\shortcite{SurfaceOnlyFerrofluids}.

In the grid-to-particle step of FLIP, we use 0.95 FLIP velocity mixed with 0.05 PIC velocity. In the particle-to-grid step we transfer not only the momentum but also the liquid's SDF. Each FLIP particle is a liquid ball, whose diameter is 1.01 times half of the voxel diagonal.

For moving FLIP domains, we update the location of the fill and sink zones at the beginning of each sub-step before particle advection. The sub-step size must be small enough so no particles moves outside the updated fill zone. All air gaps are immediately filled by particles after the advection step.
In the fill and remove particles step, we convert the BEM mesh to a SDF with functions provided by OpenVDB. When we fill new particles, we must ensure that after the new particles turned into a smoothed SDF, the smoothed FLIP SDF should coincide with the BEM SDF, so that the water level is not altered by coupling. We found generating a particle if its center is slightly ($0.7$ of the particle radius) interior to the BEM surface reproduces the same SDF after smoothing.
We remove all FLIP particles in the sink zone (domain buffer zone where the BEM SDF is greater than the -0.7 radius threshold). 
Around the fill zone (domain buffer zone with BEM mesh and SDF smaller than zero) we create a narrow band velocity volume and sample the interpolated velocity every four voxels according to Section \ref{sec:method:BEM2FLIP}. We count the particle numbers in each voxel in the fill zone. We generate new particles at random positions in a voxel if the count is less than 8, and remove extra particles if it is larger than 16.
The newly generated particles in the fill zone takes velocities linearly interpolated from the narrow band velocity volume. This approach saves the expensive boundary integral in Eq.~(\ref{eq:u_HD}).

The FLIP domain boundary velocity is also linearly interpolated from above narrow band velocity volume. In the pressure projection part we generally follow Weber et al.~\shortcite{weberMG}, except we use unsmooth aggregation to build the multigrid hierarchy algebraically and replace their V cycle by the W cycle. We apply three Red-Black Gauss-Seidel iterations before restriction and after prolongation in the W cycle. To define the restriction operator, each eight neighbor voxels at current level are averaged to become a larger voxel at the coarser level. The coarser level matrix is scaled by 0.5 according to St\"uben~\shortcite{AMGstubenreview} (see Section 6.4 therein).

\section{Results} \label{sec:results}

\begin{table*}
\caption{Overview of the different scenes presented in this paper. The resolution $\Delta x$ is stated in meter, and the time step size $\Delta t$ and per-invocation run time $T$ in seconds. We list the invocation counts in the parenthesis for each type of run time since adaptive sub-steps happen. The coupling is carried out once per frame. Eight particles occupy a single voxel when using FLIP. Machine A has 2 $\times$12-core Intel Xeon E5-2687v4@3.0GHz and NVIDIA RTX 2080Ti. Machine B has 2$\times$64-core AMD EPYC 7702@2.0GHz. Machine C has 2 $\times$16-core Intel Gold 6142@2.6GHz and RTX2080Ti. Machine D has 2$\times$24-core Intel Platinum 8260@2.4GHz and NVIDIA V100.}
\label{tab:res:runtime}
\begin{tabular}[t,width=0.9\textwidth]
{cccccccccccc} 
\hline
\bf{Scene} & \bf{Figure} & $N_\text{Tri}$ & $N_\text{Particle}$ & $\Delta x_\text{BEM}$ & $\Delta x_\text{FLIP}$&  $\Delta t$ & $T_\text{Frame}$  &  $T_\text{BEM}$ & $T_\text{FLIP}$ & $T_\text{Coupling}$ & Machine \\
\hline
Dam Break & \ref{fig:result:dambreak} &  20K & / & 0.03 & / & 1/2000 & 4.7 (4000) & 3.8 (4925)&  / & / & A\\
\hline
\multirow{5}{*}{Crown Splash} & \ref{fig:res:BEM_VS_FLIP_crownsplash:BEM} &  24K & / & 0.02 & / & 1/60 & 6.8 (600) & 5.2 (786)&  / & / &A\\
& \ref{fig:res:BEM_VS_FLIP_crownsplash:FULLHDBEM} &  33K & / & 0.02 & / & 1/60 & 8.2 (600) & 5.3 (932)&  / & / & A\\
& \ref{fig:res:BEM_VS_FLIP_crownsplash:FLIPBEM} &  20K & 5M & 0.02 & 0.01 & 1/60 & 6.3 (600) & 3.7 (717)&  0.7 (999) & 0.6  & A\\
& \ref{fig:res:BEM_VS_FLIP_crownsplash:smallFLIP} &  / & 5M & / & 0.01 & 1/60 & 2.0 (600) & / &  0.7 (1707) & / & A\\
& \ref{fig:res:BEM_VS_FLIP_crownsplash:referenceFLIP} &  / & 15M & / & 0.01 & 1/60 & 4.3 (600) & / &  1.62 (1598) & / & A\\
\hline
Boat (Straight,& \ref{fig:res:wake_pattern_depth}f &  / & 1705M & / & 0.14 & 1/24 & 70.5 (480) & / &  70.5 (480) & / &B\\
Half Resolution)& \ref{fig:res:wake_pattern_depth}g &  / & 320M & / & 0.14 & 1/24 & 21.6 (480) & / &  21.6 (480) & /  &B\\
\hline
\multirow{4}{*}{Boat (Straight)}& \ref{fig:res:wake_pattern_depth}d &  65K & 99M & 0.7 & 0.07 & 1/24 &  49.3 (480) & 10.1 (480) &  10.6 (1415) & 7.1 &C\\
& \ref{fig:res:wake_pattern_depth}h &  55K & 99M & 0.7 & 0.07 & 1/24 & 42.4 (480) & 7.4 (480) &  9.3 (1414) & 6.8 &A\\
& \ref{fig:overturning_wave_hybrid}a &  111K & 21M & 0.14 & 0.07 & 1/100 & 59.7 (500) & 55.3 (501) &  2.4 (500) & 1.8 &A\\
& \ref{fig:overturning_wave_hybrid}b &  / & 210M & / & 0.07 & 1/100 & 18.2 (500) & / &  18.2 (500) & / &A\\
\hline
Boat (Wavy) & \ref{fig:result:night_sine_boat} &  89K & 230M & 0.7 & 0.07 & 1/24 & 105.1 (480) & 15.8 (923) &  22.5 (1404) & 9.8  &C\\
\hline
Boats (Cycling) &   \ref{fig:teaser}, \ref{fig:result:triple_boat} &  170K & 634M & 0.7 & 0.07 & 1/24 & 248.9 (600) & 41.2 (600) &  64.3 (1766) & 15.8 &D\\
\hline
\end{tabular}
\end{table*}

In this section, we present examples to evaluate our hybrid method. In all experiments, we use eight particles per voxel in the FLIP simulation, and reduce the maximum relative divergence error in the pressure equation to $10^{-7}$.
First a BEM dam break example shows that the adopted BEM formulation captures 3D fluid physics. Then we use a reference FLIP simulation to drive both the shallow water equation \cite{chentanez2015coupling} and BEM, to show that the BEM can extend water waves with the correct shape and speed while SWE cannot. After we test the FLIP to BEM coupling, we show our BEM to FLIP coupling based on the boundary integral is compatible with Stokes's wave theory \cite{FAB}. Next we validate the two-way coupling by extending a small FLIP region with BEM to a larger tank, whose motion qualitatively matches the reference FLIP solution in the larger tank. Furthermore, we evaluate our method where FLIP handles liquid-solid collisions to complement the weakness of BEM. We explore the choice of parameters on the shape of the boat wake pattern, and show the wake pattern can match real world cases measured on airborne images \cite{KelvinMachrabaud2013boat}. Finally we show that our hybrid method supports multiple FLIP domains, and supports secondary white water simulation due to the horizontal velocity field provided by the BEM mesh. Relevant parameters and run times are summarized in Table~\ref{tab:res:runtime}.

\subsection{Dam Break}
\label{sec:result:dambreak}

\begin{figure}
\includegraphics[width=0.40\textwidth,trim={1cm 0 1cm 2cm},clip]{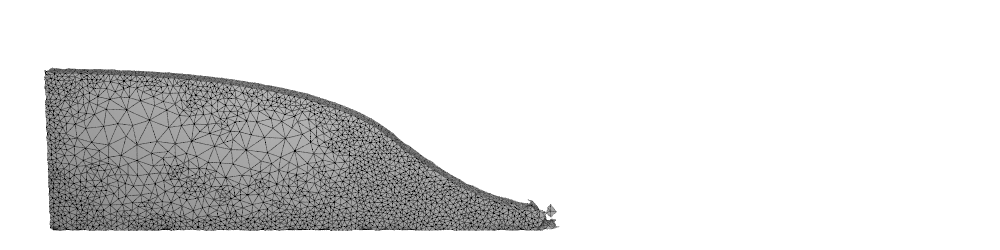}
\includegraphics[width=0.40\textwidth,trim={1cm 0 1cm 2cm},clip]{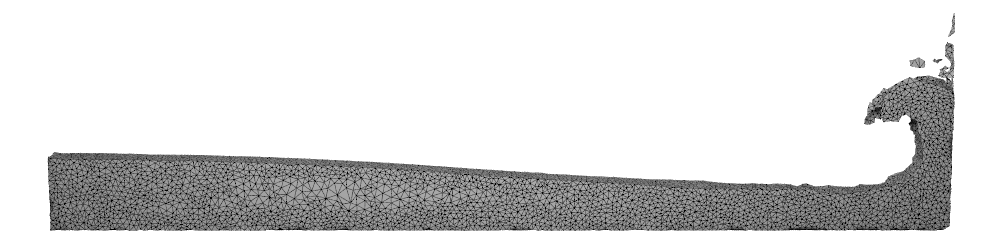}
\includegraphics[width=0.40\textwidth,trim={1cm 0 1cm 2cm},clip]{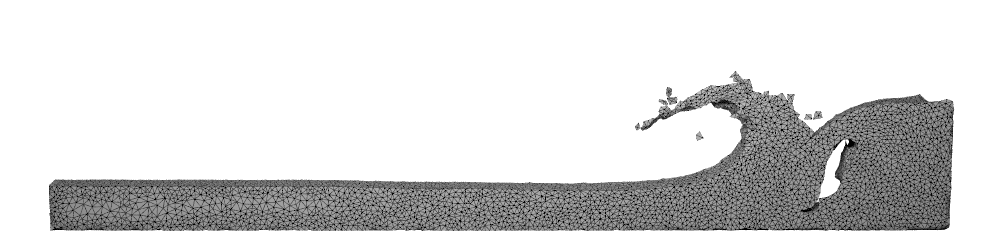}
\caption{Dam break simulated by BEM following the setup of Colagrossi and Landrini~\shortcite{SPHdambreakcolagrossi2003numerical}. Our choice for the BEM formulation can capture the defining turning waves of this experiment as well as the turbulent motions of the surface.}
\label{fig:result:dambreak}
\end{figure}

Our first experiment showcases the ability of the surface-only liquids formulation to capture 3D liquid physics. This distinguishes our BEM solver from height-field- or wave-equation-based options. We position a $2\,\textrm{m}\times1\,\textrm{m}\times0.2\,\textrm{m}$ fluid volume at the corner of a $5.366\,\textrm{m}\times3\,\textrm{m}\times0.2\,\textrm{m}$ container according to the setup of Colagrossi and Landrini~\shortcite{SPHdambreakcolagrossi2003numerical} and let it flow freely. Our choice of the BEM solver is able to qualitatively match the shape of their simulation as shown in Figure~\ref{fig:result:dambreak}.

\subsection{FLIP Influencing SWE and BEM}

\label{sec:result:FLIP_influence_SWE_BEM}
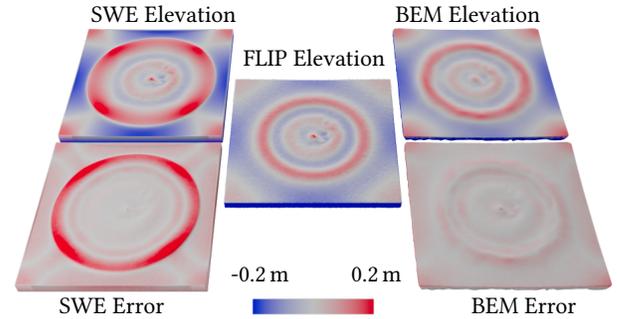
\begin{figure}
    \centering
    \begin{overpic}[width=0.47\textwidth]{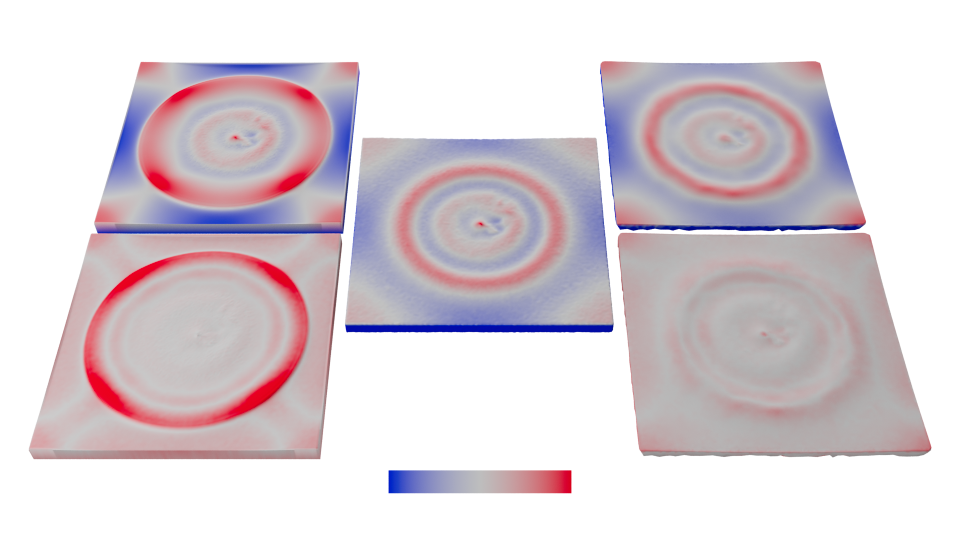}
    \put(15,51){SWE Elevation}
    \put(63,51){BEM Elevation}
    \put(39,44){FLIP Elevation}
    \put(10,5){SWE Error}
    \put(75,5){BEM Error}
    \put(37,10){-0.2\,m}
    \put(56,10){0.2\,m}
    \end{overpic}
    \caption{SWE and BEM simulations driven by a reference FLIP simulation. Middle: the reference FLIP simulation. Top left (right): The SWE (BEM) simulation driven by the reference FLIP. Blue and red indicate low and high elevation. Bottom left (right): The same SWE (BEM) simulation color encoded with nearest distance to the reference FLIP mesh. When waves propagate out of the influence region, SWE water waves have incorrect shapes and speeds while BEM waves can qualitatively match the reference.}
    \label{fig:res:SWE_FLIP_BEM}
\end{figure}

Previous methods on 2D-3D coupling extend the 3D fluid simulation with shallow water equation \cite{nilsshallowwatercouple,chentanez2015coupling}. However, shallow water equation cannot propagate the water wave with the correct shape and speed. We drop a water ball with 0.7 m radius from 3 m above the ground at the center of a $20\,\textrm{m}\times1\,\textrm{m}\times20\,\textrm{m}$ shallow tank. The SWE, FLIP and BEM simulation cover the same domain, with $\Delta x=0.1, 0.1, 0.2 \textrm{m}$ respectively. The reference FLIP simulation drives the SWE simulation within a 6 m-radius circular region at the center according to \cite{chentanez2015coupling}, and drives the BEM simulation within a square region with 6 m edge length at the center. The result is shown in Figure \ref{fig:res:SWE_FLIP_BEM}. When waves propagate out of the influence region, the SWE water waves have incorrect shapes and speeds while BEM waves can qualitatively match the reference. Note this example already use a shallow tank. The SWE would perform worse in a deeper tank.

\subsection{Boundary Integral Velocity Interpolation}
\label{sec:result:boundary_integral}
\begin{figure}
    \centering
    \begin{overpic}[width=0.47\textwidth]{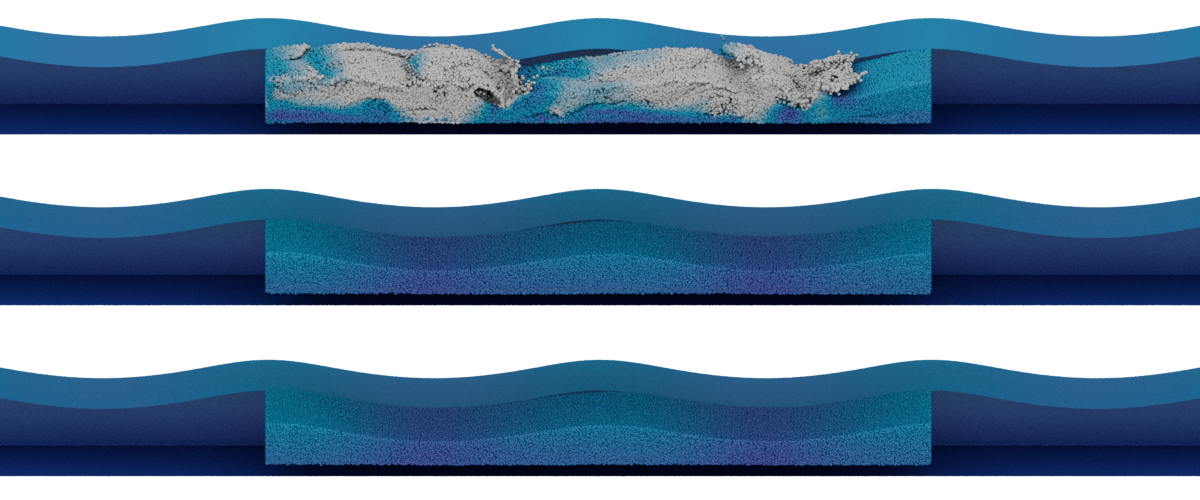}
    \put(0,41){Nearest Neighbor}
    \put(0,27){Ours}
    \put(0,13){Stokes Theory}
    \end{overpic}
    \caption{Comparison of velocity interpolation method for a Stokes wave mesh. Top: Nearest neighbor interpolation. Middle: Our boundary integral interpolation. Bottom: Reference Stokes wave theory velocity. The nearest neighbor interpolation does not provide a correct boundary condition for the FLIP to maintain Stokes's wave. Our interpolation method can produce the correct result.}
    \label{fig:res:Stokes_wave}
\end{figure}

We determine the velocity inside the BEM mesh by computing the boundary integral of the velocity on the mesh. This approach is compatible with Stokes's wave theory adopted in the FAB method \cite{FAB}. We start with a mesh that represents the shape of a Stokes wave, with correct mesh surface velocity field according to Stokes's theory both on and below the curvy free surface. The simulation is carried out in a reference frame moving along with the wave, so the mesh look static over time. We carve out a piece of mesh and fill with FLIP particles. With the correct flow boundary condition, the FLIP simulation should reproduce the same Stokes wave. In Figure \ref{fig:res:Stokes_wave} we compare the nearest neighbor interpolation, our interpolation, as well as the velocity directly from Stokes's theory. Our interpolation method is able to correctly recover the fluid velocity inside the mesh.

\subsection{Two-Way Coupling for Domain Extension}
\label{sec:res:ablation}

\begin{figure}
\begin{subfigure}[h]{{.47\textwidth}}
\includegraphics[width=0.24\textwidth,trim={0 0 0 3cm},clip]{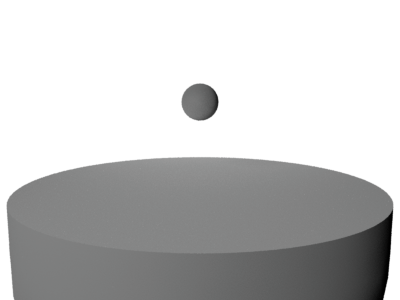}
\includegraphics[width=0.24\textwidth,trim={0 0 0 3cm},clip]{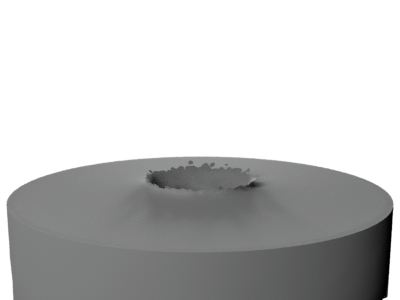}
\includegraphics[width=0.24\textwidth,trim={0 0 0 3cm},clip]{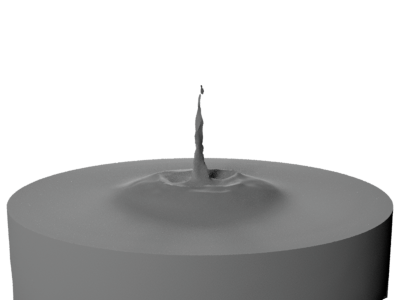}
\includegraphics[width=0.24\textwidth,trim={0 0 0 3cm},clip]{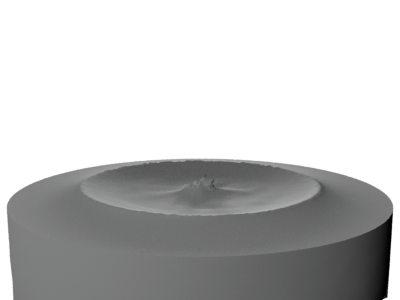}
\caption{Pure BEM \cite{SurfaceOnlyLiquids}.}
\label{fig:res:BEM_VS_FLIP_crownsplash:BEM}
\end{subfigure}
\vspace{-3pt}
\begin{subfigure}[h]{{.47\textwidth}}
\includegraphics[width=0.24\textwidth,trim={0 0 0 2cm},clip]{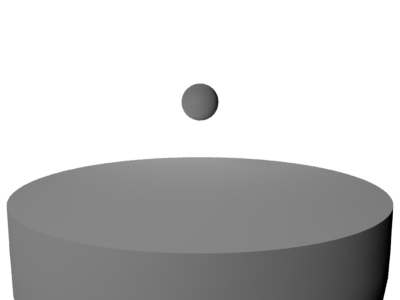}
\includegraphics[width=0.24\textwidth,trim={0 0 0 2cm},clip]{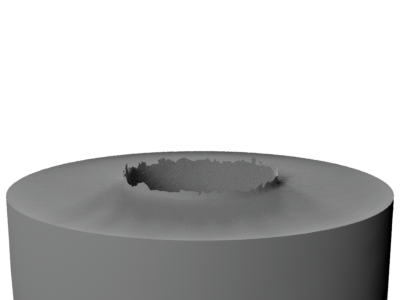}
\includegraphics[width=0.24\textwidth,trim={0 0 0 2cm},clip]{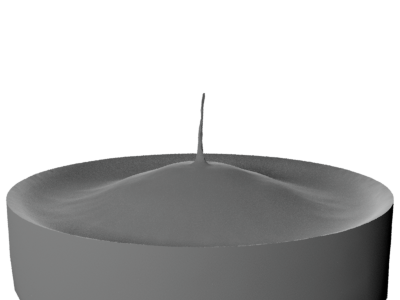}
\includegraphics[width=0.24\textwidth,trim={0 0 0 2cm},clip]{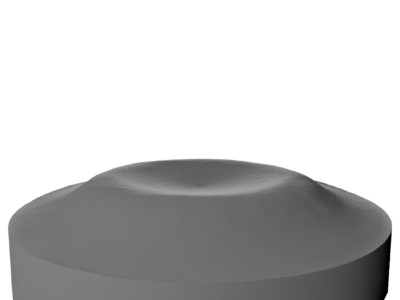}
\caption{Pure Full H.D.~BEM \cite{SurfaceOnlyFerrofluids}.}
\label{fig:res:BEM_VS_FLIP_crownsplash:FULLHDBEM}
\end{subfigure}
\vspace{-3pt}
\begin{subfigure}[h]{{.47\textwidth}}
\includegraphics[width=0.24\textwidth,trim={0 0 0 2cm},clip]{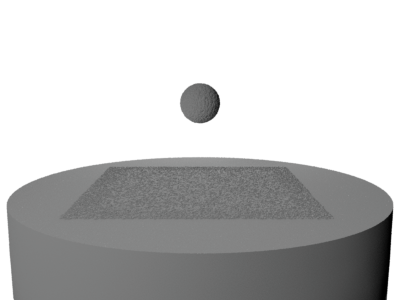}
\includegraphics[width=0.24\textwidth,trim={0 0 0 2cm},clip]{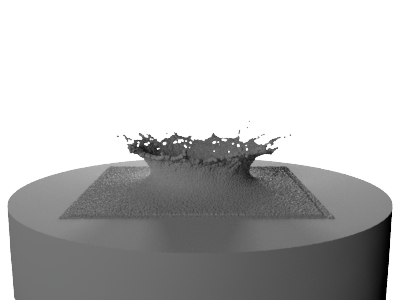}
\includegraphics[width=0.24\textwidth,trim={0 0 0 2cm},clip]{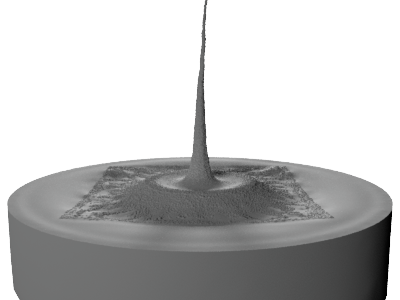}
\includegraphics[width=0.24\textwidth,trim={0 0 0 2cm},clip]{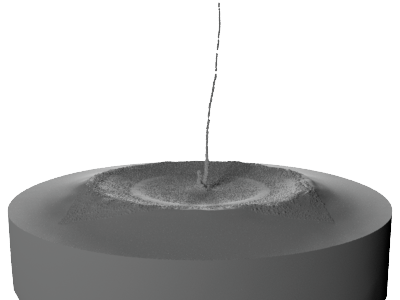}
\caption{Our hybrid approach.}
\label{fig:res:BEM_VS_FLIP_crownsplash:FLIPBEM}
\end{subfigure}
\vspace{-3pt}
\begin{subfigure}[h]{{.47\textwidth}}
\includegraphics[width=0.24\textwidth,trim={0 0 0 2cm},clip]{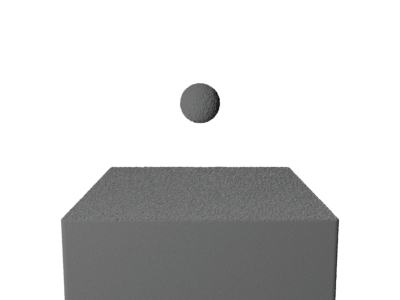}
\includegraphics[width=0.24\textwidth,trim={0 0 0 2cm},clip]{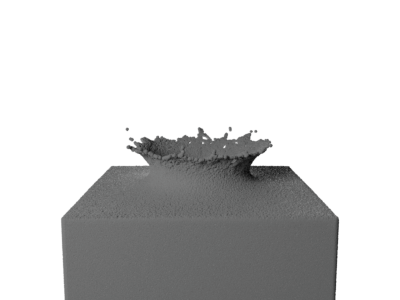}
\includegraphics[width=0.24\textwidth,trim={0 0 0 2cm},clip]{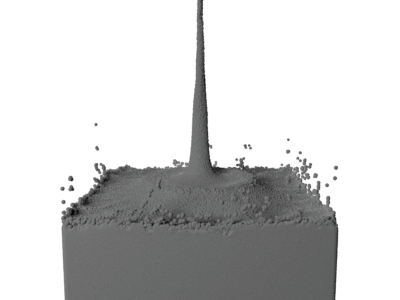}
\includegraphics[width=0.24\textwidth,trim={0 0 0 2cm},clip]{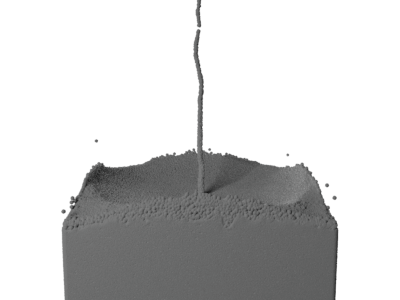}
\caption{Center FLIP without coupling.}
\label{fig:res:BEM_VS_FLIP_crownsplash:smallFLIP}
\end{subfigure}
\vspace{-3pt}
\begin{subfigure}[h]{{.47\textwidth}}
\includegraphics[width=0.24\textwidth,trim={0 0 0 2cm},clip]{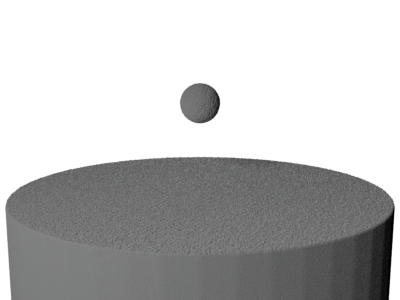}
\includegraphics[width=0.24\textwidth,trim={0 0 0 2cm},clip]{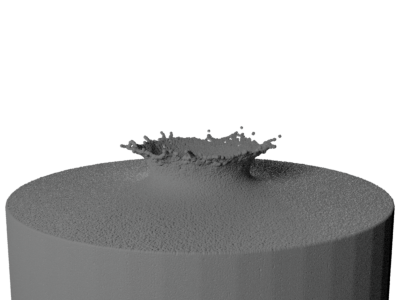}
\includegraphics[width=0.24\textwidth,trim={0 0 0 2cm},clip]{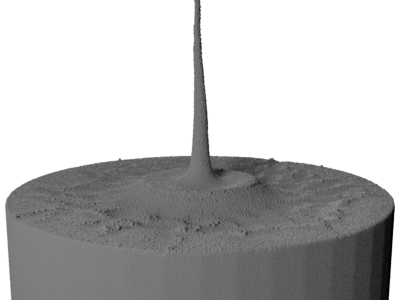}
\includegraphics[width=0.24\textwidth,trim={0 0 0 2cm},clip]{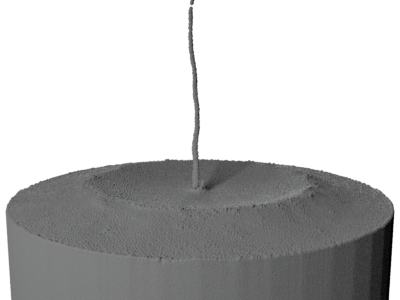}
\caption{Reference FLIP.}
\label{fig:res:BEM_VS_FLIP_crownsplash:referenceFLIP}
\end{subfigure}
\vspace{-3pt}
\caption{Simulation of a water ball dropping into a cylinder pool. The pure BEM simulation (a) and our hybrid FLIP-BEM simulation (c) with particles and BEM mesh can match the reference FLIP simulation (e), but the BEM with full H.D.~(b) does not match the reference FLIP simulation (e). The FLIP simulation without coupling (d) is affected by its boundary.}
\label{fig:res:BEM_VS_FLIP_crownsplash}
\end{figure}

We validate our two-way coupling scheme in a domain extension problem. A water ball with 0.1\,m radius impacts a $1.2\,\textrm{m}\times 0.4\,\textrm{m} \times 1.2\,\textrm{m}$ square FLIP domain from 1 m above the waterline. The waves are extended by the BEM, and touch the outer cylinder boundary with 1\,m radius and 0.6\,m depth. The reflected waves in BEM propagate back to the small FLIP domain. Ideally, the motion should match a reference FLIP simulation in the larger domain.
In this example, the minimum edge length of the triangle in the BEM simulation is $0.02\,\textrm{m}$. The voxel size of the FLIP simulation is $0.01\,\textrm{m}$.

The FLIP particles are integrated with midpoint rule with CFL condition = 1. The results are shown in Figure~\ref{fig:res:BEM_VS_FLIP_crownsplash}.

We prefer the partial H.D. (Helmholtz decomposition). formulation of Da et al.~\shortcite{SurfaceOnlyLiquids} over the full H.D. of Huang and Michels \shortcite{SurfaceOnlyFerrofluids}. We show the different H.D.~in Figure \ref{fig:res:BEM_VS_FLIP_crownsplash:BEM} and \ref{fig:res:BEM_VS_FLIP_crownsplash:FULLHDBEM} respectively. The BEM solver with partial H.D.~is able to qualitatively match the dynamic motion of the reference FLIP simulation. However, the full H.D.~leads to quite different results compared to the reference FLIP simulation. 

Since BEM alone already matches the reference FLIP qualitatively, using it to extend the small FLIP domain can probably produce a matching result as well. This is indeed the case. As shown in Figure~\ref{fig:res:BEM_VS_FLIP_crownsplash:FLIPBEM}, the hybrid simulation looks similar to the reference one. Initially, the BEM mesh is static and the momentum is transferred from the water drop in the center FLIP simulation. Our hybrid method allows the water wave to propagate beyond the FLIP box boundary and to reach the true cylinder boundary. There is visible square pattern because we render the FLIP particles with their simulation radius, so part of the sphere intersects the BEM mesh. In Figure \ref{fig:res:BEM_VS_FLIP_crownsplash:FLIPBEM} (second from left), the interior particles rise slightly higher than the transition zone. This is an artifact of our weakly coupling scheme. Here, when the water ball impacts the surface, the velocity interpolated from BEM does not generate enough outflow of the liquid on the FLIP domain boundary. Therefore, the particles inside the small FLIP domain was squeezed by the impact and rise up. A strong coupling scheme may solve this issue, but it may be much more expensive to solve the system. The BEM vertex inside the FLIP domain is governed by FLIP system with BEM velocity boundary, while the BEM vertex just outside is governed by BEM system. The seam indicates how these two systems match. The two system is different, but most of the time, their difference is small.
In contrast, as shown in Figure~\ref{fig:res:BEM_VS_FLIP_crownsplash:smallFLIP}, without the BEM coupling, the FLIP simulation alone is influenced by the box boundary leading to different fluid motion.


\subsection{Wake Pattern}
\label{sec:speedboat}

\begin{figure*}
    \centering
    \begin{overpic}[width=0.97\textwidth]{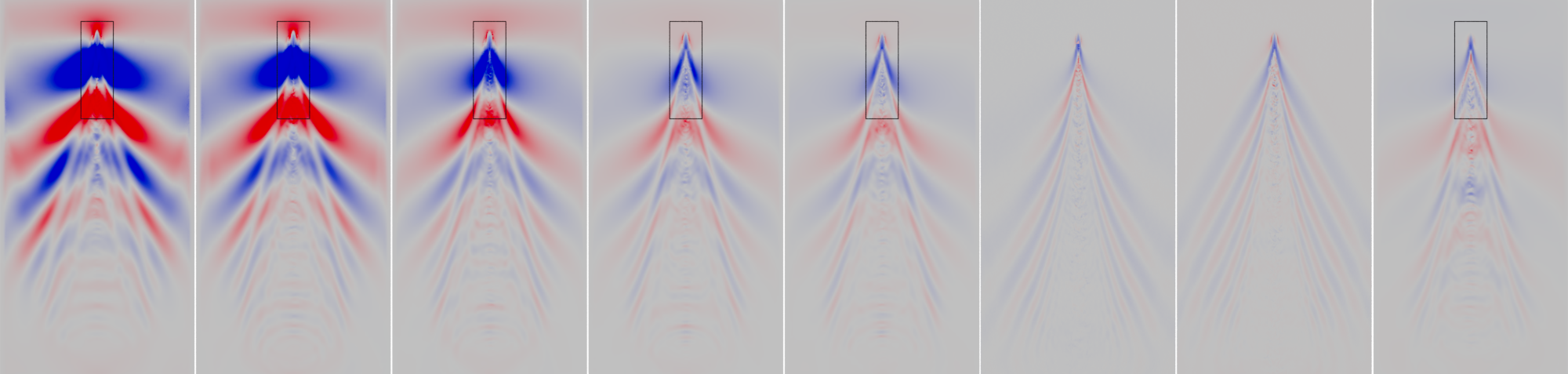}
    \put(0.5,22){(a)}
    \put(13,22){(b)}
    \put(25.5,22){(c)}
    \put(38,22){(d)}
    \put(50.5,22){(e)}
    \put(63,22){(f)}
    \put(75.5,22){(g)}
    \put(88,22){(h)}
    \end{overpic}
    \vspace{-2mm}
    \caption{Wake patterns of a sailing boat on a $120\,\textrm{m} \times 240\,\textrm{m}$ domain with different configurations. Figures a-e correspond to hybrid methods with $\Delta x_\text{FLIP}=0.07\,\textrm{m}$, $\Delta x_\text{BEM}=0.7\,\textrm{m}$, FLIP depth $3.5 \,\textrm{m}$, artificial friction coefficient = 0.8, and BEM samples FLIP velocity 0,1,2,3,4 voxels below the FLIP surface. Figure f and g correspond to FLIP simulation with $20$ and $3.5\,\textrm{m}$ depth, and $\Delta x_\text{FLIP}=0.14 \,\textrm{m}$. Figure h is without artificial friction, and samples three voxels below the interface. Blue and red colors map to -0.8\,m and 0.8\,m elevation in all figures. Black boxes indicate the FLIP regions.}
    \label{fig:res:wake_pattern_depth}
\end{figure*}
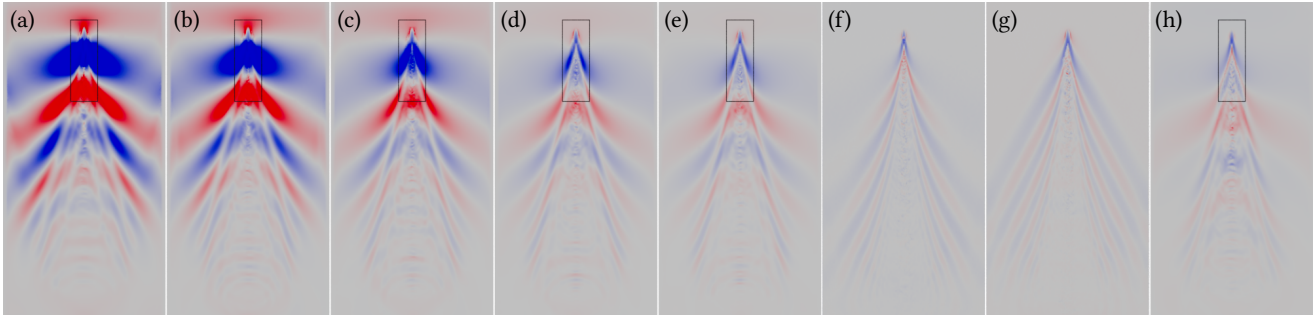
We analyze our hybrid method for simulating the ship/boat wakes. In particular, we analyze key factors influencing the wake amplitudes, compare the wake angles to real world measurements, and examine the overturning waves behind the ship/boat.

The wake amplitudes and wake angle tests are carried out in a $120\,\textrm{m} \times 20\,\textrm{m} \times 240 \,\textrm{m}$ domain. The boat is about 10\,m long. The coupled FLIP domain is $20\,\textrm{m} \times 3.5\,\textrm{m} \times 60 \,\textrm{m}$. The water flow velocity is $10\,\textrm{m/s}.$

%

The influence of sampling depth $\alpha$ in Eq.~(\ref{eq:sample_lower}) is illustrated in Figure \ref{fig:res:wake_pattern_depth}. In Figure \ref{fig:res:wake_pattern_depth}a-e, we sample at 0,1,2,3,4 voxels below the FLIP surface respectively. The artificial friction coefficient is 0.8. In Figure \ref{fig:res:wake_pattern_depth}f, we run the reference FLIP simulation, and in Figure \ref{fig:res:wake_pattern_depth}g, we only simulate a shallow 3.5 m tank to examine influence of depth on the wake pattern. Additionally, in Figure \ref{fig:res:wake_pattern_depth}h we turn off the artificial friction, and sample 3 voxels below. With increasing sampling depth, the amplitudes are closer to that of the reference FLIP. For the same sampling depth, turning off the artificial friction also leads to an amplitude closer to the reference as shown in Figure \ref{fig:res:wake_pattern_depth}h. The shallow tank simulation in Figure \ref{fig:res:wake_pattern_depth}g has a different water wave dispersion relationship compared to Figure \ref{fig:res:wake_pattern_depth}f.

\begin{figure*}
    \centering
    \begin{overpic}[width=0.98\textwidth]{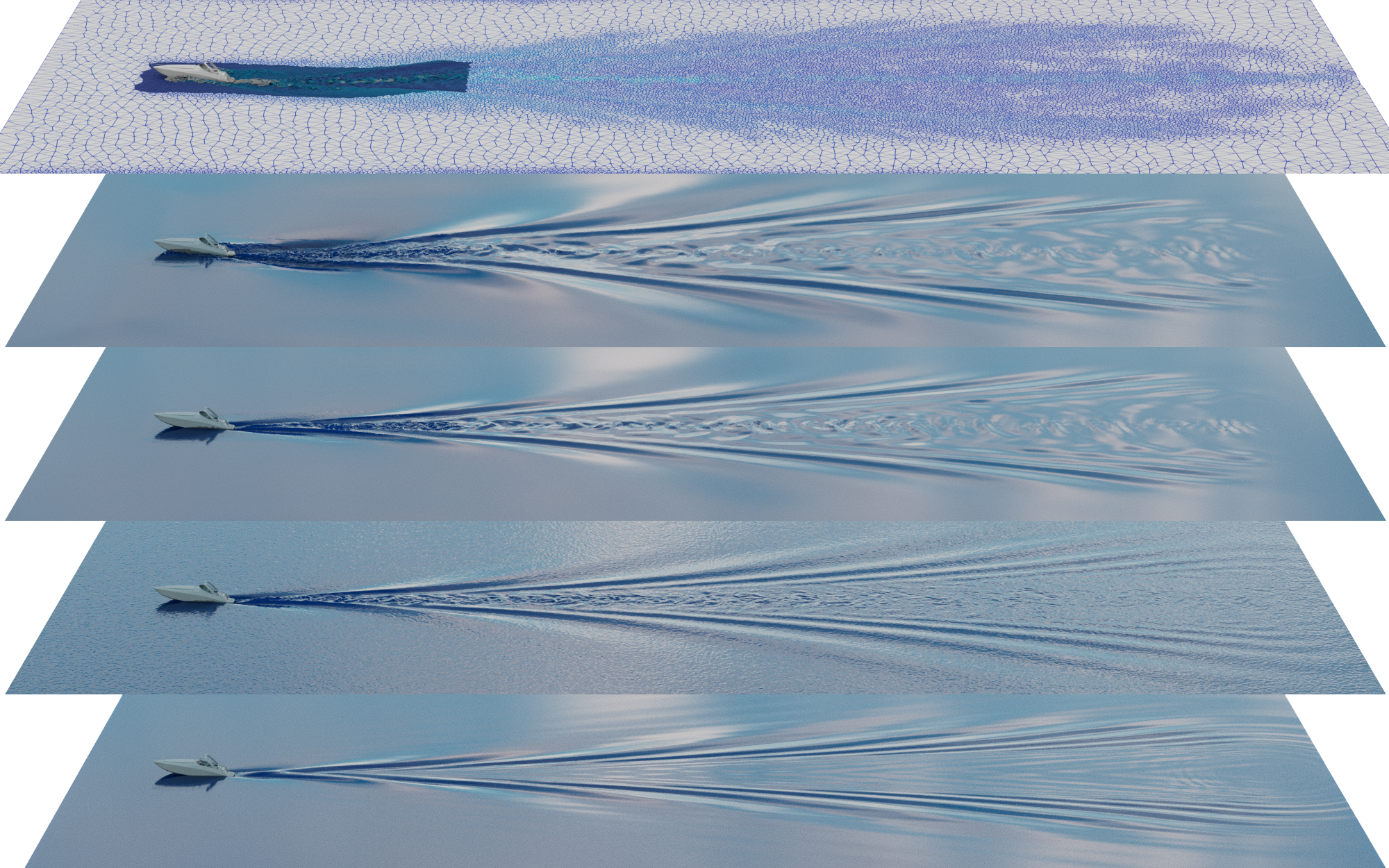}%
    \put(1,60){(a)}%
    \put(1,47.5){(b)}%
    \put(1,35){(c)}%
    \put(1,22.5){(d)}%
    \put(1,10){(e)}%
    \end{overpic}
    \caption{
    Simulated boat wakes. In realistic renderings, FLIP particles are converted to mesh, and rendered with BEM mesh. See Section \ref{sec:res:triple_boat} for detailed steps. (a) FLIP particles and BEM mesh from hybrid simulation in Figure \ref{fig:res:wake_pattern_depth}d. (b) Realistic rendering of hybrid simulation in Figure \ref{fig:res:wake_pattern_depth}d.  (c) Realistic rendering of Figure \ref{fig:res:wake_pattern_depth}h. (d) Realistic rendering of reference FLIP simulation in Figure \ref{fig:res:wake_pattern_depth}f. (e) Simulated result with \cite{dispersionwave}.}
    
    \label{fig:friction_vs_nofriction}
\end{figure*}

\begin{figure}
    \centering
    \includegraphics[width = 0.47\textwidth]{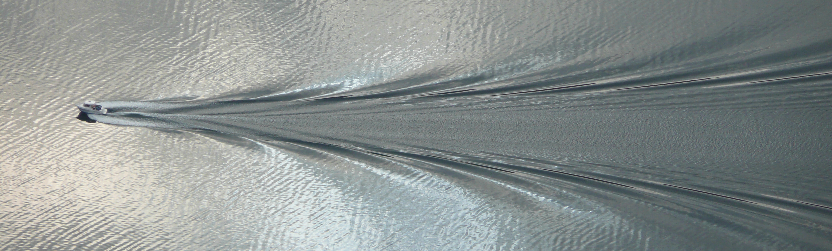}
    \caption{
     Real photo of boat wakes at  Lyse fjord in Norway, adapted from a photograph by Edmont, distributed under a CC BY-SA 3.0 license \cite{WikiWake}.
    }
   
    \label{fig:wikipediaWakes}
\end{figure}

In Figure \ref{fig:friction_vs_nofriction}, we render some wake patterns to investigate the influence of wake amplitudes on visual appearance. We choose the hybrid results with 3 voxels of sampling depth, with (Figure \ref{fig:friction_vs_nofriction}a,b) and without (Figure \ref{fig:friction_vs_nofriction}c) artificial friction. The reference FLIP simulation is presented in Figure \ref{fig:friction_vs_nofriction}d. It shows that our method, especially without artificial friction, can approximate the look of reference FLIP simulation qualitatively without the noise on FLIP surface reconstructed from particles. The reference FLIP simulation at half of the resolution of the hybrid FLIP already costs 120\,Gb memory with 1.7 billion particles. Memory consumption prevents us to compute the reference at the same resolution. A photograph of real boat wakes is in Figure \ref{fig:wikipediaWakes}. In Figure \ref{fig:friction_vs_nofriction}e, we show the result of a height field approach with correct dispersion relationship \cite{dispersionwave}.

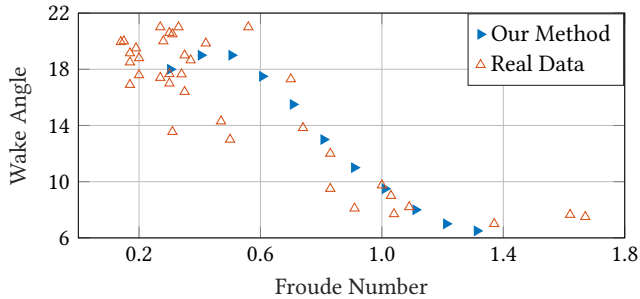
\begin{figure}[!t]
    \centering
    \definecolor{mycolor1}{rgb}{0.00000,0.44700,0.74100}%
\definecolor{mycolor2}{rgb}{0.00000,0.45098,0.74118}%
\definecolor{mycolor3}{rgb}{0.85000,0.32500,0.09800}%
\definecolor{mycolor4}{rgb}{0.92900,0.69400,0.12500}%
\definecolor{mycolor5}{rgb}{0.49400,0.18400,0.55600}%
\pgfplotsset{scaled y ticks=false}
\pgfplotsset{minor grid style={dashed}}
\pgfplotsset{
compat=1.11,
legend image code/.code={
\draw[mark repeat=2,mark phase=2]
plot coordinates {
(0cm,0cm)
(0.0cm,0cm)        
(0.3cm,0cm)         
};}}
\begin{tikzpicture}

\begin{axis}[%
width=0.85\linewidth,
height=0.35\linewidth,
scale only axis,
xmin=0,
xmax=1.8,
xlabel style={font=\color{white!15!black}},
xlabel={Froude Number},
xtick={0.2, 0.6,  1.0,  1.4,  1.8},
xticklabels={0.2,  0.6,  1.0,  1.4,  1.8},
ymin=6,
ymax=22,
ylabel style={font=\color{white!15!black}},
ylabel={Wake Angle},
ytick={6, 10, 14, 18, 22},
yticklabels={6, 10, 14, 18, 22},
axis background/.style={fill=white},
title style={font=\bfseries},
xmajorgrids,
ymajorgrids,
legend style={at={(0.999,0.999)}, anchor=north east, legend cell align=left, align=left, draw=white!15!black}
]
\addplot [color=mycolor1, only marks, mark=triangle*, mark options={solid, rotate=270, fill=mycolor2, mycolor1}]
  table[row sep=crcr]{%
   0.303045763365663	18\\
0.404061017820884	19\\
0.505076272276105	19\\
0.606091526731327	17.5000000000000\\
0.707106781186548	15.5000000000000\\
0.808122035641769	13\\
0.909137290096990	11\\
1.01015254455221	9.50000000000000\\
1.11116779900743	8\\
1.21218305346265	7\\
1.31319830791787	6.50000000000000\\
};
\addlegendentry{Our Method}

\addplot [color=mycolor3, only marks, mark=triangle, mark options={solid, mycolor3}]
  table[row sep=crcr]{%
1.37000000000000	7\\
1.67000000000000	7.50000000000000\\
1.62000000000000	7.65000000000000\\
1.04000000000000	7.70000000000000\\
0.910000000000000	8.10000000000000\\
1.09000000000000	8.20000000000000\\
1.03000000000000	9\\
0.830000000000000	9.50000000000000\\
1	9.75000000000000\\
0.830000000000000	12\\
0.500000000000000	13\\
0.310000000000000	13.5500000000000\\
0.740000000000000	13.8300000000000\\
0.470000000000000	14.3000000000000\\
0.350000000000000	16.4000000000000\\
0.170000000000000	16.9000000000000\\
0.300000000000000	17\\
0.700000000000000	17.3000000000000\\
0.270000000000000	17.4000000000000\\
0.200000000000000	17.5900000000000\\
0.340000000000000	17.6500000000000\\
0.300000000000000	17.6700000000000\\
0.170000000000000	18.5000000000000\\
0.370000000000000	18.6500000000000\\
0.200000000000000	18.8000000000000\\
0.350000000000000	19\\
0.170000000000000	19.1500000000000\\
0.190000000000000	19.5000000000000\\
0.420000000000000	19.8500000000000\\
0.140000000000000	19.9500000000000\\
0.280000000000000	20\\
0.150000000000000	20\\
0.310000000000000	20.5000000000000\\
0.300000000000000	20.6000000000000\\
0.270000000000000	21\\
0.330000000000000	21\\
0.560000000000000	21\\
};
\addlegendentry{Real Data}

\end{axis}

\end{tikzpicture}%
    \vspace{-12pt}
    \caption{Wake angle as a function of the hull Froude number $Fr = U/\sqrt{gL}$, where $U$ is the flow velocity, $g$ is the gravity, and $L$ is the length of the boat. Our results match the measurements based on airborne images from \cite{KelvinMachrabaud2013boat}.}
    
    \label{fig:res:Angle_VS_Froude}
\end{figure}

Next we analyze the wake angles. The wakes behind a ship/boat are usually referred to as the Kelvin wakes, which are characterized by a fixed angle around $20^\circ$ and feather like pattern. The fixed angle is due to the dispersion relationship of deep water waves of all wavelengths. However in reality,  Rabaud and Moisy \shortcite{KelvinMachrabaud2013boat} reported the transition of fixed Kelvin angle to increasingly smaller Mach angles when the ship/boat moves faster. They believe "the finite size of the disturbance (ship) explains this transition". Our hybrid approach reproduces their discoveries, matching their theory as well as their real world measurements from airborne images. In Figure \ref{fig:res:Angle_VS_Froude} we can observe the transition with increasing speed.

\begin{figure}
    \centering
    \includegraphics[width=0.15\textwidth]{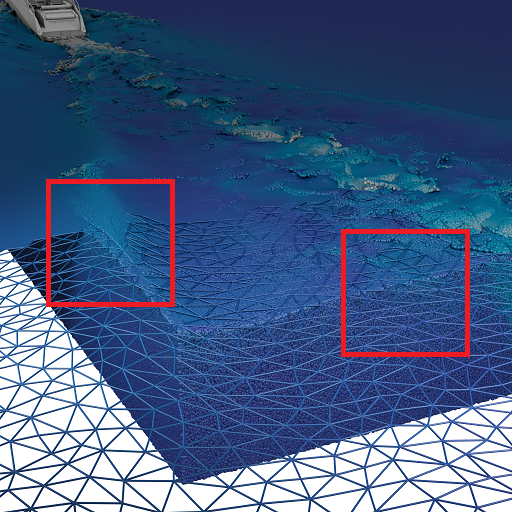}
    \includegraphics[width=0.15\textwidth]{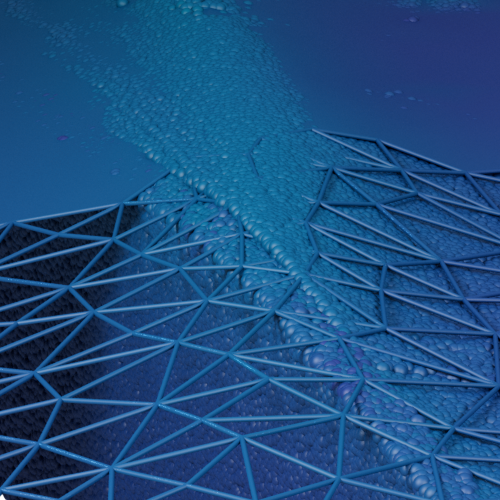}
    \includegraphics[width=0.15\textwidth]{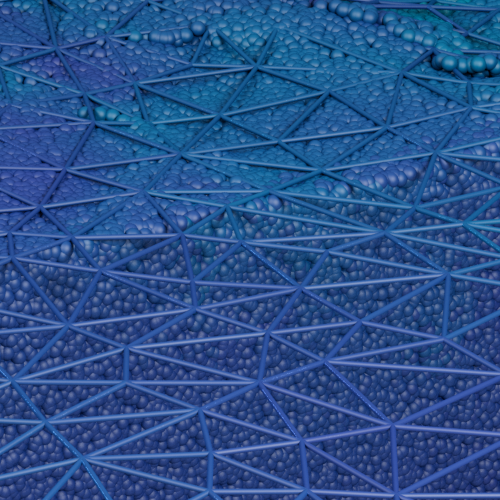}
    \caption{FLIP-BEM transition boundary in the hybrid simulation in Figure \ref{fig:res:wake_pattern_depth}d. The transition zone is visible in the first zoomed in figure, since particles above the BEM mesh are removed in the coupling algorithm to match the BEM free surface. Particles and triangles are colored by velocity.}
    \label{fig:FLIPBEM_transition}
\end{figure}

In the transition zone between FLIP and BEM we fill and remove FLIP particles to track the BEM free surface. This is visualized in Figure \ref{fig:FLIPBEM_transition}.

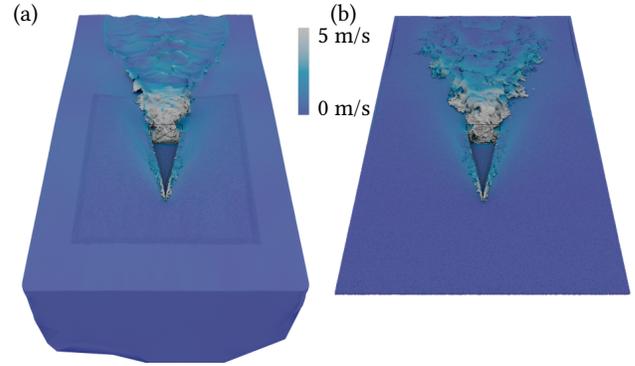
\begin{figure}
    \centering
    \begin{overpic}[width = 0.47\textwidth]{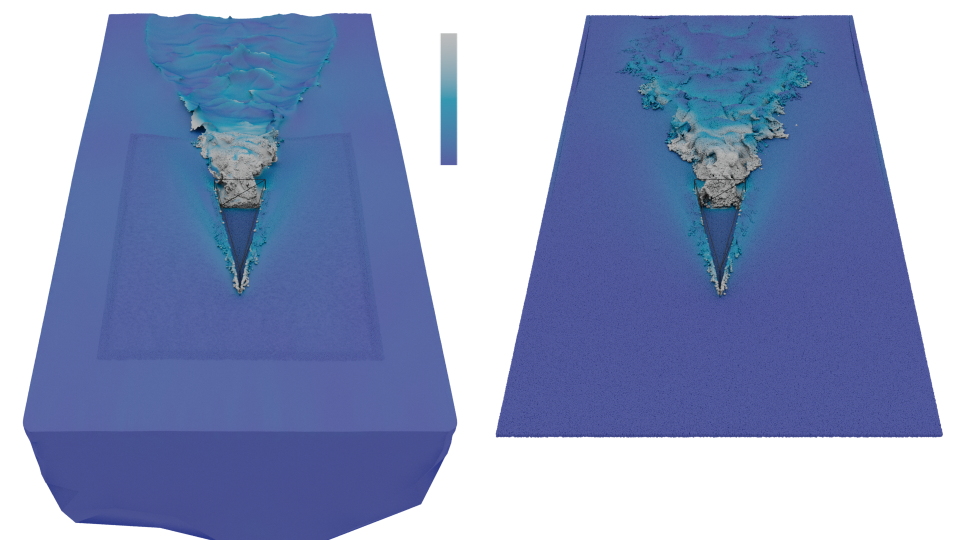}
    \put(1,54){(a)}
    \put(51,54){(b)}
    \put(49,50.5){5 m/s}
    \put(49,39) {0 m/s}
    \end{overpic}
    \caption{
    Extending overturning waves behind a wedge with (a) particles and the BEM mesh in our hybrid method and (b) particles in the reference FLIP simulation. Only the top layer of particles are exported to save IO. The domain size is $20\,\textrm{m}\times 10\,\textrm{m}\times{45}\,\textrm{m}$. $\Delta x_\text{BEM}=0.14\,\textrm{m}, \Delta x_\text{FLIP} = 0.07\,\textrm{m}$
    }
    
    \label{fig:overturning_wave_hybrid}
\end{figure}

Our method is capable of extending overturning waves behind the obstacle. However, the BEM is not the best choice for such turbulent waves because it handles complex topology changes less efficiently compared to the FLIP method. In addition, the BEM is not designed for capturing vortices near the surface. Hence extending overturning waves are not the recommended usage of our approach. Nevertheless, we show the overturning wave results in Figure \ref{fig:overturning_wave_hybrid}.

\begin{figure*}
\includegraphics[width=0.196\textwidth]{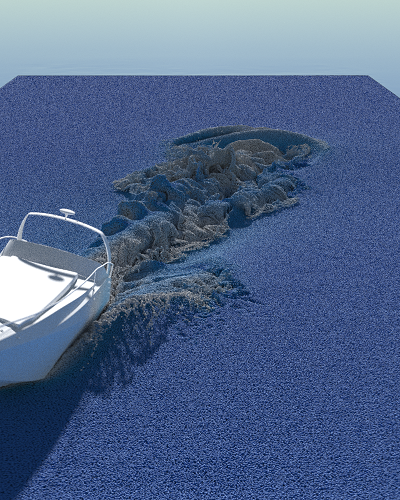}
\includegraphics[width=0.196\textwidth]{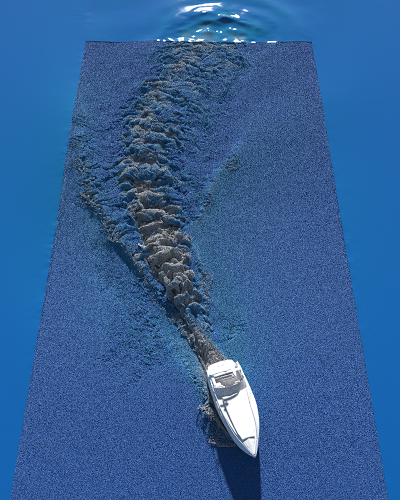}
\includegraphics[width=0.196\textwidth]{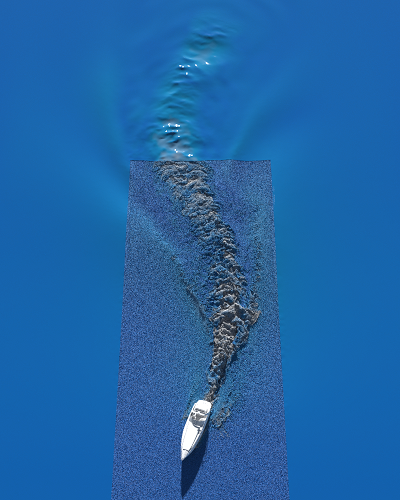}
\includegraphics[width=0.196\textwidth]{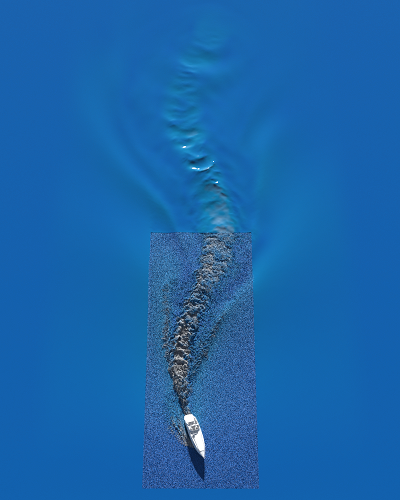}
\includegraphics[width=0.196\textwidth]{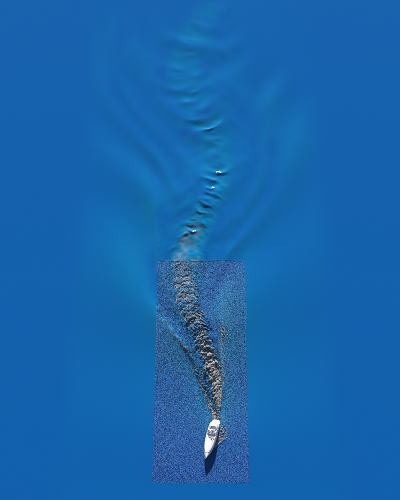}
\includegraphics[width=0.196\textwidth]{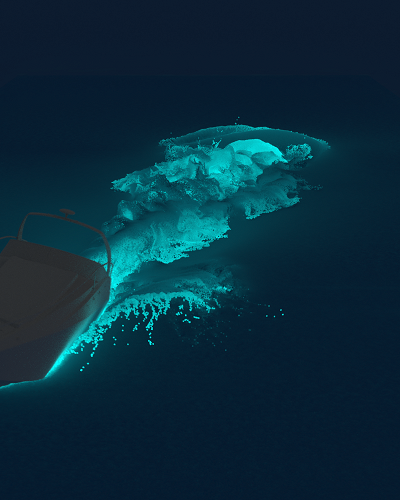}
\includegraphics[width=0.196\textwidth]{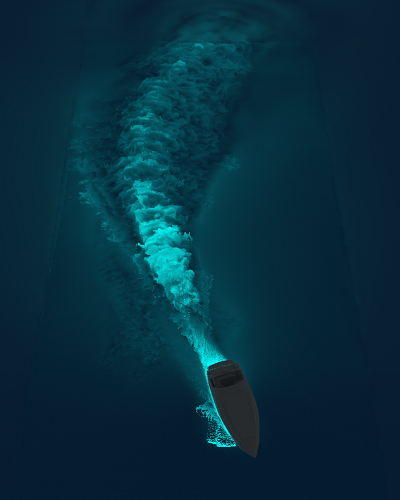}
\includegraphics[width=0.196\textwidth]{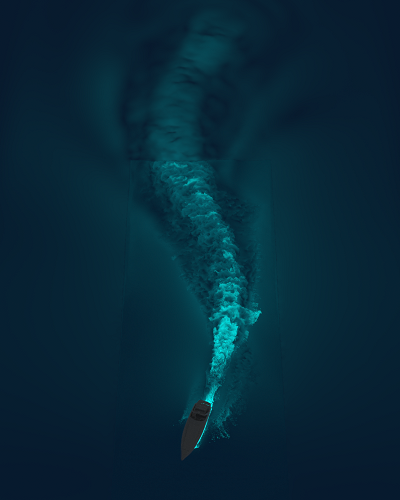}
\includegraphics[width=0.196\textwidth]{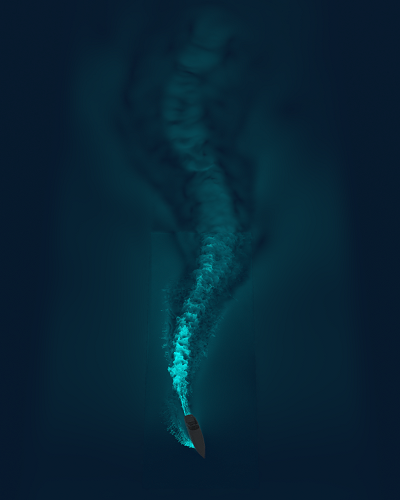}
\includegraphics[width=0.196\textwidth]{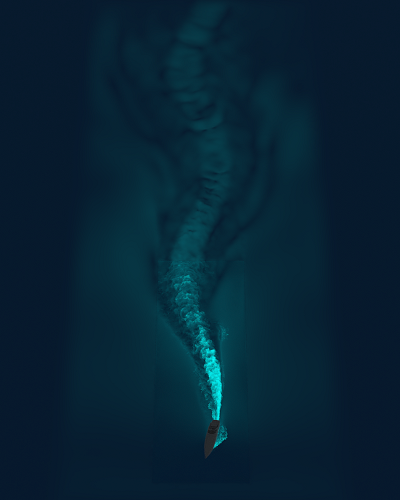}\\
\vspace{-2mm}
\caption{Simulation of a speedboat following a serpentine movement. The BEM allows for the natural extension of the waves. Particles are rendered inside the FLIP domain. The BEM mesh subtracting the FLIP domain patch are rendered on the outside.}
\label{fig:result:night_sine_boat}
\end{figure*}

In a more complex example, the boat moves like a sine wave. The FLIP domain size is $32\,\textrm{m}\times 3.5\,\textrm{m}\times 85\,\textrm{m}$. The results are shown in Figure~\ref{fig:result:night_sine_boat}.

\subsection{Multiple Moving Domains and Rendering}
\label{sec:res:triple_boat}
\begin{figure*}
\begin{overpic}[width=1.0\textwidth]{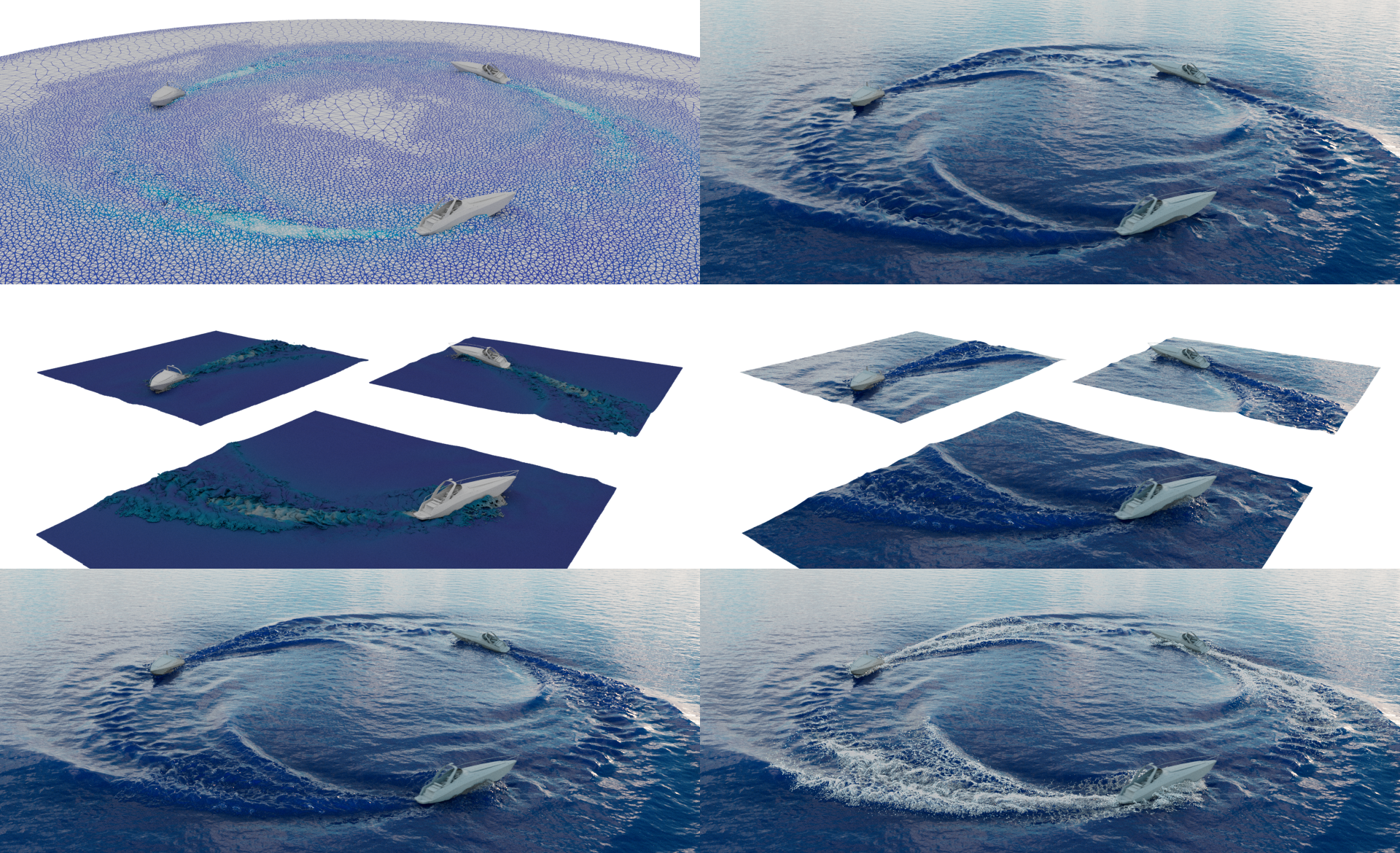}
\put(0.5,59){(a)}
\put(50.5,59){(b)}
\put(0.5,38.7){(c)}
\put(50.5,38.7){(d)}
\put(0.5,18.4){(e)}
\put(50.5,18.4){(f)}
\end{overpic}
\vspace{-6mm}
\caption{Simulation of speedboats riding in a circle computed with our hybrid method. (a) Raw BEM mesh without subdivision colored by velocity. (b) Rendered BEM mesh and exterior ocean with FFT waves. (c) Raw top layer FLIP particles colored by velocity. (d) Rendered FLIP mesh with FFT waves. (e) Composed image of b and d. (f) Composed image of a secondary whitewater simulation. The water surface is perturbed by the previous FLIP domain and affects the following FLIP domain.}
\label{fig:result:triple_boat}
\end{figure*}

In this example, we show that the method handles multiple moving FLIP domains interacting with each other through BEM meshes. As shown in Figure~\ref{fig:teaser}, and \ref{fig:result:triple_boat}, three FLIP domains follow the cycling motion of three boats. The boat at the front excites the water surface. When a FLIP domain in the back sweeps through the perturbed region, newly generated particles follow the perturbed water surface and its motion. There are three $50\,\textrm{m}\times 3.5\,\textrm{m}\times 50\,\textrm{m}$ FLIP domains in a cylinder tank with 100 m radius and 20 m depth. The track radius is 40 m while speedboats travel at 10 m/s. In principle, the FLIP domains can overlap, because they only interact with each other through the BEM mesh. However, we avoid doing so because it can introduce extra difficulty in rendering overlapping surfaces.

We use this scene as an example to demonstrate a simplified production pipeline. Our hybrid method allows secondary whitewater simulation to enhance the final look. This is possible because the BEM mesh provides surface velocity field. We convert the BEM mesh to SDF with OpenVDB, and create narrow band volumetric velocity field around the mesh according to the closest point velocity on the BEM mesh for each voxel. The BEM SDF and velocity field are replaced by FLIP SDFs and velocity fields inside FLIP domains. Such combined velocity field and SDF are used for the off-the-shelf whitewater simulation included in Houdini \cite{Houdini}.

The final image is composed of different render passes. Initially we have a BEM mesh (Figure\ref{fig:result:triple_boat}a). We smooth it and flatten the edge near BEM boundary to match a flat ocean surface. The intersection of BEM mesh and a large cylinder is rendered together with the exterior flat plane subtracting the same cylinder to create seamless ocean. Next, FFT waves \cite{jerryocean} are added at low velocity regions. This forms the first pass (Figure \ref{fig:result:triple_boat}b).

We only output the surface layer of particles (Figure \ref{fig:result:triple_boat}c) and the other particles are exported as a SDF at simulation resolution to save disk consumption. The surface particles are turned into a higher resolution SDF and merged with the rest simulation resolution SDF. The combined SDF are turned into polygons. We extract the FLIP surface by taking the intersection of three FLIP polygon meshes and three boxes slightly smaller than the FLIP domains. In order to reduce artifacts on the FLIP-BEM boundary, near the intersecting box, we pull the intersected FLIP mesh vertices towards the nearest point on smoothed BEM mesh. The closer they are to the intersecting box, the more they cling to the smoothed BEM mesh. These flattened FLIP meshes are augmented with FFT waves to render the high resolution liquid mesh (Figure \ref{fig:result:triple_boat}d). When rendering Figure \ref{fig:result:triple_boat}d, the smoothed BEM mesh subtracting the three boxes participates in light transport, but are not rendered (Houdini phantom).

Figure \ref{fig:result:triple_boat}d is put on top of Figure \ref{fig:result:triple_boat}b to get the rendered liquid surface in Figure \ref{fig:result:triple_boat}e. Finally the whitewater pass are added in Figure \ref{fig:result:triple_boat}f.




\section{Runtime Performance}

\begin{table}
\caption{Runtime comparison of our FLIP and Houdini \cite{Houdini}, which implements narrow band FLIP (NB) \cite{narrowband} and adaptive octree solver for pressure (Octree) \cite{RyanOctree}. The comparison is on the same machine A in Table \ref{tab:res:runtime} with three substeps/frame. The first two rows corresponds to the FLIP region in Figure \ref{fig:res:wake_pattern_depth}h. The last two rows corresponds to brute force simulation of high resolution ($\Delta x_\text{FLIP}=0.07$\,m) shallow ocean in Figure \ref{fig:res:wake_pattern_depth}g ($\Delta x_\text{FLIP}=0.14$\,m). Our FLIP outperforms Houdini even with stricter accuracy ($\varepsilon$) for the pressure solver.}

\label{tab:comparinghoudini}
\scalebox{0.8}{
\begin{tabular}[t,width=0.47\textwidth]
{ccccccc} 
\hline
\bf{Method} & \bf{NB} & \bf{Octree} & \bf{DOF} & \bf{Liquid Voxels} & $\varepsilon$ & \bf{t/frame}\\
\hline
Our FLIP & NO &  NO & 12.2M & $285\times50\times857$ & $10^{-7}$ & 28s\\
\hline
\multirow{3}{*}{Houdini} & YES &  YES & 2.4M & $285\times50\times857$ & $10^{-4}$ & 43s\\
& YES &  NO & 12.2M & $285\times50\times857$ & $10^{-4}$ & 90s\\
& NO &  NO & 12.2M & $285\times50\times857$ & $10^{-4}$ & 105s\\
\hline
Our FLIP& NO &  NO & 293M & $1714\times50\times3429$ & $10^{-7}$ & 576s\\
\hline
Houdini & YES &  YES & 50M & $1714\times50\times3429$ & $10^{-4}$ & 1077s\\
\hline
\end{tabular}}
\end{table}

One of the motivations of our work is the limited runtime performance of existing methods when simulating a large body of water. In this section we compare to several other possible methods in terms of runtime performance. The runtime performances of some methods are only available for different scenes and machine configurations. Therefore we scale these runtimes linearly for rough estimation.

First we compare variations of FLIP solvers. We compare our FLIP with narrow band FLIP \cite{narrowband} and adaptive octree pressure solver \cite{RyanOctree} provided by Houdini \cite{Houdini} directly on the same machine. The comparison is summarized in Table \ref{tab:comparinghoudini}. First we compare the FLIP part in our hybrid simulation of wake pattern (Figure \ref{fig:res:wake_pattern_depth}). Our FLIP implementation is faster than the two improvements combined. Furthermore we try to simulate the whole ocean with brute force. However, neither Houdini nor our FLIP can fit into the memory. Therefore we simulate the first three sub-steps of the shallow ocean in Figure \ref{fig:res:wake_pattern_depth} but at high resolution, which amounts to extending the FLIP region horizontally to fit the BEM region. Tall-cells methods \cite{tallcell,chentanez2011real} are slower than uniform methods for the same degrees of freedom (DOF). Even if they run as fast as the uniform methods with the same DOF, given the runtime in Table \ref{tab:comparinghoudini}, a 5-cell-thick simulation of the large domain with tall-cells methods would take 58s , still slower than our hybid method (42s in Table \ref{tab:res:runtime}).

Next we roughly estimate the performance of a recent octree simulator \cite{ryoichi2020adaptive}. It provides not only adaptivity inside the liquid volume, but also on the liquid surface, making it very suitable for both deep and wide liquid simulation. Our hybrid framework is comparable to its high performance. They reported 20s/substep for a water tank scene with equivalently 33.3M uniform liquid voxels (or 266M particles) on a 8-core Intel CPU at 3.6 GHz.

We estimate the run time if we use such octree simulator for the FLIP part of our straight boat wake scene in Figure \ref{fig:res:wake_pattern_depth}. Taking into account of CPU frequency and core numbers, their simulator would take $20\,\textrm{s/substep}\times 8\, \textrm{cores}/24\,\textrm{cores}\times 3.6\,\textrm{GHz}/3.0\,\textrm{GHz}\times99\textrm{M}/266\textrm{M}=3.0\,\textrm{s/substep}$. Hence compared to our 9.3 s/substep FLIP performance in Table \ref{tab:res:runtime}, it is three times faster. The BEM domain is 24 times larger and 7 times deeper. It is possible to simulate this using the octree method because these additional volumes would be filled with coarse grids. However, whether it can preserve wakes and simulate the whole domain efficiently cannot be concluded with our estimation. In their airplane scene, the wakes seem to dissipate quickly.

\section{Discussion and Future Work} \label{sec:discussion}

In this paper, we hybridize a local high-resolution FLIP solver and a surface-based BEM solver for multiscale ocean modeling. This method allows us to naturally extend the FLIP domain to a larger extent as shown in the crown splash experiment (Figure~\ref{fig:res:BEM_VS_FLIP_crownsplash}). The accuracy is demonstrated by comparing the boat wake pattern to the brute-force FLIP results and a real example (Figure~\ref{fig:res:wake_pattern_depth}, \ref{fig:friction_vs_nofriction} \ref{fig:res:Angle_VS_Froude}). Through the BEM mesh, multiple volumetric domains can interact with each other and collectively generate complex patterns in a large body of water (Figure~\ref{fig:result:triple_boat}). Thanks to the surface representation of liquid and adaptive triangle sizes of the BEM, the BEM mesh can cheaply expand both horizontally and vertically.
 
Our method is suitable for open water simulation with both water splashes near the main object, and waves that travel long distances. The most time consuming part of the simulation is the volumetric solver near the main object, even with our optimized implementation. The extra time induced by the BEM simulation in the calm region and coupling is usually no more than $30\%$. The FLIP part consumes about $90\%$ of the memory, up to 120\,Gb for the 1.7 billon particles scene. It is up to the user to decide the resolution and size of the FLIP region in the BEM water body depending on the machine capability. The BEM is limited by the acceleration grid for collision detection. Currently, it uses uniform grids with a cell size equal to the smallest triangle edge length. Hence, the bounding box of the BEM cannot be too large.

 The performance of the BEM is the main drawback. It is limited by the serial surface tracking program LosTopos \cite{LosTopos}, which takes quite a long time to resolve collisions and topology changes, etc. The BEM involves dense matrix-vector multiplication as a result of integral equation systems. In principle, the fast multipole method (FMM) \cite{greengard1987fast} can solve the matrix-vector multiplication problem, but for the problem sizes in our experiments, the naive summation on the GPU is faster given the overhead of the FMM. In our most challenging triple boat scene, the BEM simulation takes up to 2 minutes per frame for 170K triangles. About half of the time is spent on remeshing. Therefore, the smart usage of the BEM is for calm waves, rather than splashes and overturning waves. When the main object is placed in a stormy ocean, the propagated waves are negligible. The prescribed waves with one-way interaction are more efficient. The BEM does not interacts with beaches, cliffs either, because of the irregularity of solids unsupported by LosTopos.
 Another issue is the wake amplitudes in Figure \ref{fig:res:wake_pattern_depth}. We alleviate this issue by sampling below the interface. We believe this problem is due to the Helmholtz decomposition which interprets slow velocity as resistance. The resistance causes the elevation of the BEM mesh ahead of the boat. When the detailed water waves in the FLIP domain enter the BEM domain, most of the details will be lost because the BEM mesh has lower resolution, dedicated for large wave features.

 The coupling workflow may look heuristic compared to a strong coupling framework that bridges a dense integral equation system in BEM and a sparse equation system in FLIP. However, at this stage we are satisfied with its results, considering potentially larger implementation effort of a strong coupling system, and corresponding lower efficiency of a mixed integral and differential equation system.

For the problem of extending waves with correct dispersion relationships, we believe the dispersion kernel method \cite{dispersionwave} shows great potential. We look forward to seeing coupling of such height field methods with correct dispersion relationships with 3D simulations. Canabal et al.~directly modified the height field, so the height field does not have horizontal displacement or velocity information. The two-way coupling for domain extension may not be trivial due to their inherently different physics.

It would be interesting to see our method applied to other types of hybrid Lagrangian-Eulerian simulations such as IQ-MPM \cite{IQMPM}.

\section*{Acknowledgements}
\label{sec:acknowledgements}

This work has been supported by KAUST (individual baseline funding of the Computational Sciences Group within the Visual Computing Center), the National Science Foundation (NSF CAREER IIS-1943199), and CCF-1813624 and ECCS-2023780.

We thank Da Fang and Christopher Batty for releasing their source code and Miguelangelo Rosario for providing the ship model. The valuable comments of the anonymous reviewers are gratefully acknowledged.

\bibliographystyle{ACM-Reference-Format}
\bibliography{references}

\end{document}